\documentclass[twocolumn]{aastex63}

\shorttitle{Black Hole UCXBs as Galactic low-frequency GW sources}
\shortauthors{Qin et al.}

\begin{document}


\title{Black Hole Ultracompact X-Ray Binaries as Galactic Low-frequency Gravitational Wave Sources: the He Star Channel}


\author[0000-0001-9206-1641]{Ke Qin}
 \affil{School of Science, Qingdao University of Technology, Qingdao 266525, People's Republic of China; chenwc@pku.edu.cn}
\affil{School of Physics, Zhengzhou University, Zhengzhou 450001, People's Republic of China}
\author[0000-0002-9739-8929]{Kun Xu}
 \affil{School of Science, Qingdao University of Technology, Qingdao 266525, People's Republic of China; chenwc@pku.edu.cn}
\author[0000-0002-3007-8197]{Dong-Dong Liu}
  \affil{Key Laboratory for the Structure and Evolution of Celestial Objects, Yunnan Observatories, Chinese Academy of Sciences, Kunming 650216, People's Republic of China}
\author[0000-0002-2479-1295]{Long Jiang}
  \affil{School of Science, Qingdao University of Technology, Qingdao 266525, People's Republic of China; chenwc@pku.edu.cn}
   \affil{School of Physics and Electrical Information, Shangqiu Normal University, Shangqiu 476000, People's Republic of China}
\author[0000-0002-3231-1167]{Bo Wang}
  \affil{Key Laboratory for the Structure and Evolution of Celestial Objects, Yunnan Observatories, Chinese Academy of Sciences, Kunming 650216, People's Republic of China}
\author[0000-0002-0785-5349]{Wen-Cong Chen}
  \affil{School of Science, Qingdao University of Technology, Qingdao 266525, People's Republic of China; chenwc@pku.edu.cn}
  \affil{School of Physics and Electrical Information, Shangqiu Normal University, Shangqiu 476000, People's Republic of China}



\begin{abstract}
Black hole (BH) ultracompact X-ray binaries (UCXBs) are potential Galactic low-frequency gravitational wave (GW) sources. As an alternative channel, BH UCXBs can evolve from BH+He star binaries. In this work, we perform a detailed stellar evolution model for the formation and evolution of BH UCXBs evolving from the He star channel to diagnose their detectability as low-frequency GW sources. Our calculations found that some nascent BH+He star binaries after the common-envelope (CE) phase could evolve into UCXB-LISA sources with a maximum GW frequency of $\sim5~\rm mHz$, which can be detected in a distance of 10 kpc (or 100 kpc). Once BH+He star systems become UCXBs through mass transfer, they would emit X-ray luminosities of $\sim10^{38}~\rm erg\, s^{-1}$, making them ideal multimessenger objects. If the initial He-star masses are $\geq 0.7 M_{\odot}$, those systems are likely to experience two Roche lobe overflows, and the X-ray luminosity can reach a maximum of $3.5\times 10^{39}~\rm erg\, s^{-1}$ in the second mass-transfer stage. The initial He-star masses and initial orbital periods of progenitors of Galactic BH UCXB-LISA sources are in the range of 0.32-2.9 $M_{\odot}$ and 0.02-0.19 days, respectively. Nearly all BH+He star binaries in the above parameter space can evolve into GW sources whose chirp masses can be accurately measured. Employing a population synthesis simulation, we predict the birthrate and detection number of Galactic BH UCXB-LISA source evolving from the He star channel are $R=2.2\times10^{-6}~\rm yr^{-1}$ and 33 for an optimistic CE parameter, respectively.
\end{abstract}

\keywords{binaries: close -- galaxies: helium -- stars: black holes -- star: evolution -- gravitational waves}

\section{Introduction}
Ultracompact X-ray binaries (UCXBs) are low-mass X-ray binaries (LMXBs) with ultra-short orbital periods (usually less than 60 minutes), consisting of an accreting compact object and a hydrogen-poor donor star \citep{nels86,nele10a}. UCXBs can help us to understand the angular momentum loss mechanisms \citep{ma09,haaf12a,haaf12b}, the common-envelope (CE) evolution \citep{zhu12}, and the accretion process of compact objects \citep{lin18}, thus they are ideal laboratories for testing stellar and binary evolutionary theory. Furthermore, UCXBs can emit continuous low-frequency gravitational wave (GW) signals, which could be detected by the space-borne GW detectors such as the Laser Interferometer Space Antenna \citep[LISA;][]{amar17,amar23}, TianQin \citep{luo16,wang19,huan20}, and Taiji \citep{ruan20}. Therefore, UCXBs are ideal objects pursuing multi-messenger investigations \citep{chen20}.

Thus far, the number of confirmed UCXBs (with an accurately measured orbital period $\leq 80$ minutes) and candidates is around 45 \citep{arma23}. Based on the accurate detection of orbital periods, 20 sources were identified to be UCXBs in high confidence, which includes 11 persistent sources and 9 transient sources \citep{int07,liu07,hein13,cart13,piet19,coti21,peng21,arma23}.
Among confirmed UCXBs, 19 sources were discovered to include a neutron star (NS). At present, there exist two black hole (BH) UCXB candidates. The first one is the luminous X-ray source X9 in globular cluster 47 Tucanae, which was thought to be a BH accreting from a white dwarf (WD) in a close orbit \citep{mill15,bahr17,chur17,tudo18}. Recently, \cite{mout23} found some evidences that the compact object in UCXB 4U 0614+091 may be a BH by the simultaneous NICER and NuSTAR observations. \cite{chen20} estimated that the main-sequence (MS) channel can form 60-80 NS UCXB-LISA sources in the Galaxy. However, LISA can only detect $\sim4$ BH UCXBs evolving from the MS channel \citep{qin23}. If the ratio of the numbers between NS and BH UCXB-LISA sources is similar to that of the identified numbers between NS and BH UCXBs, the number of confirmed BH UCXBs evolving from the MS channel is $\sim1$, which is comparable to the present observation. Therefore, it is difficult to observe BH UCXBs.

Because of the compactness of UCXBs, their donor stars are generally thought to be partially or completely degenerate stars such as WDs or helium (He) stars \citep{rapp82,pods02,delo03}. Optical spectroscopic analysis of UCXBs can help us to identify the properties of donor stars \citep{nele04,nele06}. Employed X-ray, ultraviolet, and optical spectroscopy, chemical elements of accretion disks in some UCXBs may include He, C, N, O, Ne, and Si \citep{nele10b}. Such a diversity of chemical compositions implied those donor stars should evolve to different nuclear-burning stages and interior degeneracy, which requires different evolutionary models to account for the formation of UCXBs \citep{seng17}. It is generally thought that UCXBs in the Galactic field evolved from the following three channels: the WD channel, the evolved MS star channel, and the He star channel \citep{post06,nele10b}.

In the first channel, the progenitors of UCXBs are compact binaries consisting of a NS/BH and a low-mass WD, in which GW radiation drives mass transfer \citep{belc04,haaf12a,jian17,yu21}. \cite{seng17} performed complete numerical models for the formation of UCXBs evolved from a stable mass transfer from a WD to an accreting NS, and found that the WD channel can reproduce the observed properties of some UCXBs with high He abundances. Using an improved mass transfer hydrodynamics model, \cite{bobr17} argued that only NS+He WD binaries with a donor-star mass less than 0.2 $M_{\odot}$ could form UCXBs through a stable mass transfer. \cite{haaf12a} also found that the donor-star masses have to be less than 0.4 $M_{\odot}$ to form UCXBs from NS+CO WD binaries.

The second channel originates from a NS/BH accreting mass from a MS donor star that fills its Roche lobe. If the MS star starts mass transfer late enough and the orbital-angular-momentum loss by magnetic braking is efficient, the system would evolve toward a UCXB. In this channel, UCXBs generally evolve from BH/NS+MS binaries whose initial orbital periods are shorter than the bifurcation period \citep{sluy05,seng17,chen20,qin23}. In the UCXB stage, the donor star is most likely to evolve into a WD. It is noteworthy that mass transfer never ceases in the whole evolutionary stage except for those systems with a NS and a fine-tuning initial orbital period \citep{chen20,qin23}.

In the He star channel, the direct progenitors of UCXBs are BH/NS+He star binaries. It is generally believed that BH/NS+He star binaries are the evolutionary products of high-mass X-ray binaries, in which the hydrogen envelope of the He-star progenitors are fully ejected in the CE stage \citep{quas19,abdu20,gotb20,wang21}. Due to the close orbit, the GW radiation dominates the orbital evolution of the nascent BH/NS+He star system and triggers a mass transfer. The BH/NS accretes He-rich materials from the He star that fills the Roche lobe and the system appears as an UCXB \citep{savo86,dewi02,hein13,wang21}.  Employing detailed stellar evolution models, \cite{jian21,jian23} found that NS/BH+He star binaries can evolve into double NS or BH+NS systems through stripped supernova explosions, and these double NS and BH+NS systems would evolve toward high-frequency GW events that can be discovered by aLIGO. Binary population synthesis (BPS) simulations indicated that binary systems consisting of a NS/BH and a naked He star can account for $\sim50\%-80\%$ UCXBs \citep{zhu12}. By a population-synthesis simulation, \cite{lomm05} proposed there are $\sim200$ BH+He star binaries and $\sim540$ NS+He star binaries in the Milky Way.

Compared with those dim WDs, He stars are close to MS stars in the Hertzsprung-Russell (H-R) diagram \citep{graf02,geie10,gotb18}. Therefore, BH+He star systems can be observed in the detached stage, and provide more evolutionary details such as the mass-transfer efficiency \citep{pack81,mink07}, and the stellar wind \citep{puls08,smit14,vink17}. In general, low-mass He stars are referred to as the hot subdwarfs \citep{han07}, while high-mass He stars are called Wolf-Rayet stars \citep{graf02}. At present, several sources consisting of a compact star and a He star have been reported. For example, LS V+22 25 (LB-1) was proposed to contain a stellar mass ($\approx 8 M_{\odot}$) BH and a low-mass (0.5-1.7 $M_{\odot}$) stripped He star \citep{eldr20,yung20}. PG 1432+159, HE 0532-4503, PG 1232-136, and PG 1743+477 were also thought to be candidates containing a BH and a He star with a very thin hydrogen envelope \citep{geie10}. However, no X-ray emissions from these detached binaries were detected. HD49798 \citep{mere09}, M101 ULX-1 \citep{liu13}, IC 10 X-1, and NGC 300 X-1 are most likely X-ray sources including He stars, in which the X-ray emissions originated from a NS/BH accreting from the stellar winds of He stars. Especially for the latter two sources, \cite{tutu16} proposed that the accretion disk and the limited X-ray luminosity of $10^{38}~\rm erg\, s^{-1}$ can be interpreted by a stellar-mass BH accreting from a Wolf-Rayet companion through the wind-Roche lobe overflow (RLOF) mechanism \citep[see also][]{mell19a,mell19b}. However, both the stellar wind accretion and the RLOF models can explain the observations of galactic strong X-ray sources Cygnus X-3 and SS 433 \citep{lomm05}, which are also thought to consist of a compact object and a mass-transferring He star. As a consequence, binary systems consisting of a compact object and a He star are potential galactic strong X-ray sources. Furthermore, they are also promising low-frequency GW sources in the Galaxy \citep{wang21,liu23}. Therefore, it is of great significance to study the evolution of BH binaries including He stars.

In this paper, we perform a detailed stellar evolution model for a large number of BH+He star binaries to investigate whether or not they can evolve toward UCXBs and low-frequency GW sources in the Milky Way. In Section 2, we describe the binary evolution code. Some detailed simulated results are shown in Section 3. The discussion and conclusion are presented in Sections 4, and 5, respectively.

\section{Binary evolution model}
The Modules for Experiments in Stellar Astrophysics \citep[MESA;][]{paxt11,paxt13,paxt15,paxt18,paxt19} is a popular code in the stellar and binary evolution field. In this work, we use a binary updated version (r12115) of the MESA to model the formation and evolution of BH UCXBs. We assume that the CE stage would produce many detached BH-He star binaries, which are taken to be the evolutionary beginning point of detailed stellar evolution. For simplicity, the BH is considered a point mass with an initial mass of $M_{\rm BH,i}=8~M_{\odot}$. The code only models the nuclear synthesis of the He star and the orbital evolution of the binary. Therefore, the evolutionary fates of BH-He star binaries depend on the initial He-star mass ($M_{\rm He,i}$) and the initial orbital period ($P_{\rm orb,i}$) for a given input physics. We build a zero age main sequence He star model with solar
metallicity \citep{wong19}, which consists of 98\% helium and 2\% metallicity (i.e. $Y=0.98, Z = 0.02$). The lowest He-star mass
is taken to be 0.32 $M_{\odot}$, under which the center He burning would extinguish \citep{iben90,han02,yung08}. For each binary system, we run the MESA code until the time step reaches a minimum time-step limit or the stellar age is greater than the Hubble time (14 Gyr).

For the wind setting of the He star, the "Dutch" options with a scaling factor of 0.8 are used in the schemes including $hot_{-}wind_{-}scheme$, $cool_{-}wind_{-}RGB_{-}scheme$, and $cool_{-}wind_{-}AGB_{-}scheme$ \citep{gleb09}. We use Type 2 opacities for extra C/O burning during and after He burning. Furthermore, the time step options with $mesh_{-}delta_{-}coeff=1.0$ and $varcontrol_{-}target=10^{-3}$ are adopted. Our inlists are available at doi:10.5281/zenodo.10075413.

The fast wind of the He star is thought to carry away its specific orbital-angular momentum. The wind-accretion efficiency of the BH via Bondi-Hoyle-like accretion is low \citep{taur17}. For example, the wind-accretion efficiency is $\sim0.003$ for a binary with a $8.8~M_{\odot}$ BH and a $6.0~M_{\odot}$ He star \citep{jian23}. Therefore, we ignore the wind accretion in the whole evolutionary process. Once the He star fills its Roche lobe, the mass transfer initiates from the donor star to the BH at a rate of $\dot{M}_{\rm tr}$. During the mass transfer, we adopt the accretion efficiency scheme given by \cite{taur06}, i.e. $\alpha= 0, \beta = 0.5$, and $\delta= 0$, here $\alpha$, $\beta$, and $\delta$ represent the fractions of mass loss from the He star in the form of fast wind, the ejected mass from the vicinity of the BH and from a circumbinary co-planar toroid, respectively. Therefore, the accretion rate of the BH is $\dot{M}_{\rm acc}=(1-\beta)\dot{M}_{\rm tr}=0.5~\dot{M}_{\rm tr}$.

The mass-growth rate of the BH is limited by the Eddington accretion rate as
\begin{equation}
\dot{M}_{\rm Edd}=\frac{4\pi GM_{\rm BH}}{\kappa c\eta},
\end{equation}
where $G$ is the gravitational constant, $c$ is the speed of light in vacuo, $M_{\rm BH}$ is the mass of the BH, $\kappa = 0.2(1+X)$ is the Thompson-scattering opacity of electrons ($X$ is the hydrogen abundance of the transferred material, and $X = 0$ for a He donor star); $\eta = 1-\sqrt{1-(M_{\rm BH}/3M_{\rm BH,0})^{2}}$ (for $M_{\rm BH}\le \sqrt{6}M_{\rm BH,0}$, $M_{\rm BH,0}$ is the initial mass of the BH) is the energy conversion efficiency of the accreting BH \citep{pods03}. Therefore, the mass-growth rate of the accreting BH is
$\dot{M}_{\rm BH}={\rm min}(0.5~\dot{M}_{\rm tr},\dot{M}_{\rm Edd})$. The excess materials in unit time ($\dot{M}_{\rm tr}-\dot{M}_{\rm BH}$) are thought to be ejected at the vicinity of the BH in the form of isotropic winds, carrying away the specific orbital angular momentum of the BH.  Thus, the angular-momentum-loss rate due to isotropic winds can be written as
\begin{equation}
\dot{J}_{\rm iso}= -\frac{2\pi a^{2}M_{\rm He}^{2}(\dot{M}_{\rm tr}-\dot{M}_{\rm BH})}{(M_{\rm BH}+M_{\rm He})^{2}P_{\rm orb}},
\end{equation}
where $a$ is the orbital separation of the binary, $M_{\rm He}$ is the mass of the He star, and $P_{\rm orb}$ is the orbital period. In our model, orbital angular momentum loss via GW radiation and mass loss (fast wind and isotropic wind) are included.

During the inspiral of two components in BH binaries, the change of the mass quadrupole produces low-frequency GW signals with a frequency of $f_{\rm gw} = 2/P_{\rm orb}$. When the systems evolve into a close orbit, the emitting GW signals may be detected by space-borne GW detectors such as LISA. For a 4 years LISA mission, the GW characteristic strain of our simulated BH binaries is given by \citep{chwc20}
\begin{equation}
h_{\rm c}\approx 3.75\times 10^{-20}\left(\frac{f_{\rm gw}}{0.001~\rm Hz}\right)^{7/6}\left(\frac{\mathcal{M}}{1~M_{\odot}}\right)^{5/3}\left(\frac{10~\rm kpc}{d}\right).
\end{equation}
$d$ is the distance of the GW sources. For simplicity, the chirp mass $\mathcal{M}$ can be expressed as
\begin{equation}
\mathcal{M}=\frac{(M_{\rm BH}M_{\rm He})^{3/5}}{(M_{\rm BH}+M_{\rm He})^{1/5}}.
\end{equation}
In the numerical calculation, the corresponding binaries are thought to be LISA sources if the calculated characteristic strain exceeds the LISA sensitivity curve given by \cite{robs19}.

It is worth noting that the chirp mass in Equation (4) should be applied in a detached system that its orbital decay is entirely caused by GW radiation. For BH UCXBs, mass transfer is inevitable to influence their orbital evolution. However, the mass transfer in compact BH binaries is driven by GW radiation and goes along a timescale close to that of GW radiation \citep{haaf12a,haaf12b}.
Therefore, the estimation of the chirp mass in Equation (4) remains reliable in semi-detached BH binaries.

\section{Simulation result}
For a fixed initial BH mass and a given input physics, the evolutionary fates of BH+He star binaries depend on $P_{\rm orb,i}$ and $M_{\rm He,i}$. As evolutionary examples, we model the evolution of 10 BH+He star binaries with different $M_{\rm He,i}$ and $P_{\rm orb,i}$, which divide into three groups as follows: (1) group 1 with $M_{\rm He,i}$ = 0.6 $M_{\odot}$, $P_{\rm orb,i}$ = 0.03, 0.06, 0.09, 0.11 days; (2) group 2 with $M_{\rm He,i}$ = 0.32, 0.4, 0.6, 0.8 $M_{\odot}$, $P_{\rm orb,i}$ = 0.06 days. (3) group 3 with $M_{\rm He,i} = 1.2, 1.8, 2.8~M_{\odot}$, $P_{\rm orb,i} = 0.06$ days. Groups 1 and 2 only experience one RLOF, while there exist two or more RLOFs for group 3. Some main evolutionary parameters of three groups are listed in Table 1.

\subsection{Orbital evolution}
Figure 1 shows the evolution of orbital periods with the stellar age for BH+He star binaries in three groups. Because of small donor-star masses, the donor stars in groups 1 and 2 fill their Roche lobes until their orbital periods are in the range of 13 to 48 minutes, thus these systems appear as ultracompact detached binaries in a long timescale. In group 1, those systems with $P_{\rm orb,i}=0.03,0.06,0.09~\rm days$ can firstly be detected by LISA as low-frequency GW sources at a distance of 10 kpc, and 100 kpc (these two distances are the typical distances of the sources in the Galaxy and the Large/Small Magellanic Cloud)(the system with $P_{\rm orb,i}=0.03~\rm days$ is visible by LISA at a distance of 10 kpc at the beginning of binary evolution), then experience RLOF and become UCXBs. The system with $P_{\rm orb,i}=0.11~\rm days$ initiates the mass transfer after it is visible by LISA at a distance of 10 kpc, then it appears a UCXB that can be detected by LISA at a distance of 10 kpc, and 100 kpc (i.e. UCXB-LISA source).

In group 2, the system with $M_{\rm He,i}=0.8~M_{\odot}$ is visible by LISA in a distance of 10 kpc at the beginning of binary evolution, then it fills its Roche lobe and appears as a UCXB-LISA source. Subsequently, continuous orbital shrinkage makes it a strong GW source that can be detected by LISA at a distance of 100 kpc. The other three systems with low donor-star masses first appear as low-frequency GW sources that can be detected by LISA at a distance of 10 kpc and 100 kpc, then evolve into UCXB-LISA sources. Because of the short orbital periods, the GW radiation drives the orbital periods of the systems in groups 1 and 2 to continuously decrease to a minimum of $6-12.4~\rm minutes$, which are very similar to the minimum orbital periods ($8-10~\rm minutes$) obtained by \cite{wang21} in NS+He star systems. Most evolutionary tracks of groups 1 and 2 emerge "knee" features, which are consistent with the positions that the RLOF starts. This phenomenon originates from the orbital expansion caused by the mass transfer from the less massive He star to the more massive BH, which dilutes the orbital decay due to GW radiation.

In group 3, three systems first appear as low-frequency GW sources that can be detected by LISA at a distance of 10 kpc, then begin case BA mass transfer at the first solid circles. The orbital period of the system with $M_{\rm He,i}=1.2~M_{\odot}$ continuously decreases after the mass transfer. However, the orbits of the other two systems with $M_{\rm He,i}=1.8$, and $2.8~M_{\odot}$ show a widening tendency. In theory, the orbit should widen when the mass is transferred from the less massive donor star to the more massive BH. In contrast, GW radiation causes the orbit to shrink. The evolutionary fate of the orbit would depend on the competition between the mass transfer and GW radiation. The orbital shrinkage of the system with $M_{\rm He,i}=1.2~M_{\odot}$ originates from a low mass-transfer rate (see also Figure 2), which can not conquer the orbital decay caused by GW radiation. After the case BA mass transfer ceases, three systems start a case BB mass transfer soon (the solid circles are approximately in the same positions as those open circles, and the timescales between these two circles are 0.346, 0.389 and 0.340 Myr for $M_{\rm He,i}=1.2, 1.8$, and 2.8 $M_{\odot}$, respectively). Subsequently, two systems $M_{\rm He,i}=1.8$, and $2.8~M_{\odot}$ can not be detected by LISA at a distance of 10 kpc due to a further expansion of the orbits. Once case BB mass transfer ends, the continuous orbital shrinkage by GW radiation causes the system with $M_{\rm He,i}=1.8~M_{\odot}$ to evolve toward LISA sources that can be detected at a distance of 10 kpc, and 100 kpc. The evolution of the systems with $M_{\rm He,i}=1,2$, and $2.8~M_{\odot}$ stops due to a numerical difficulty after the case BB mass transfer ceases.

\begin{figure}
\centering
\includegraphics[width=1.0\linewidth,trim={0 0 0 0},clip]{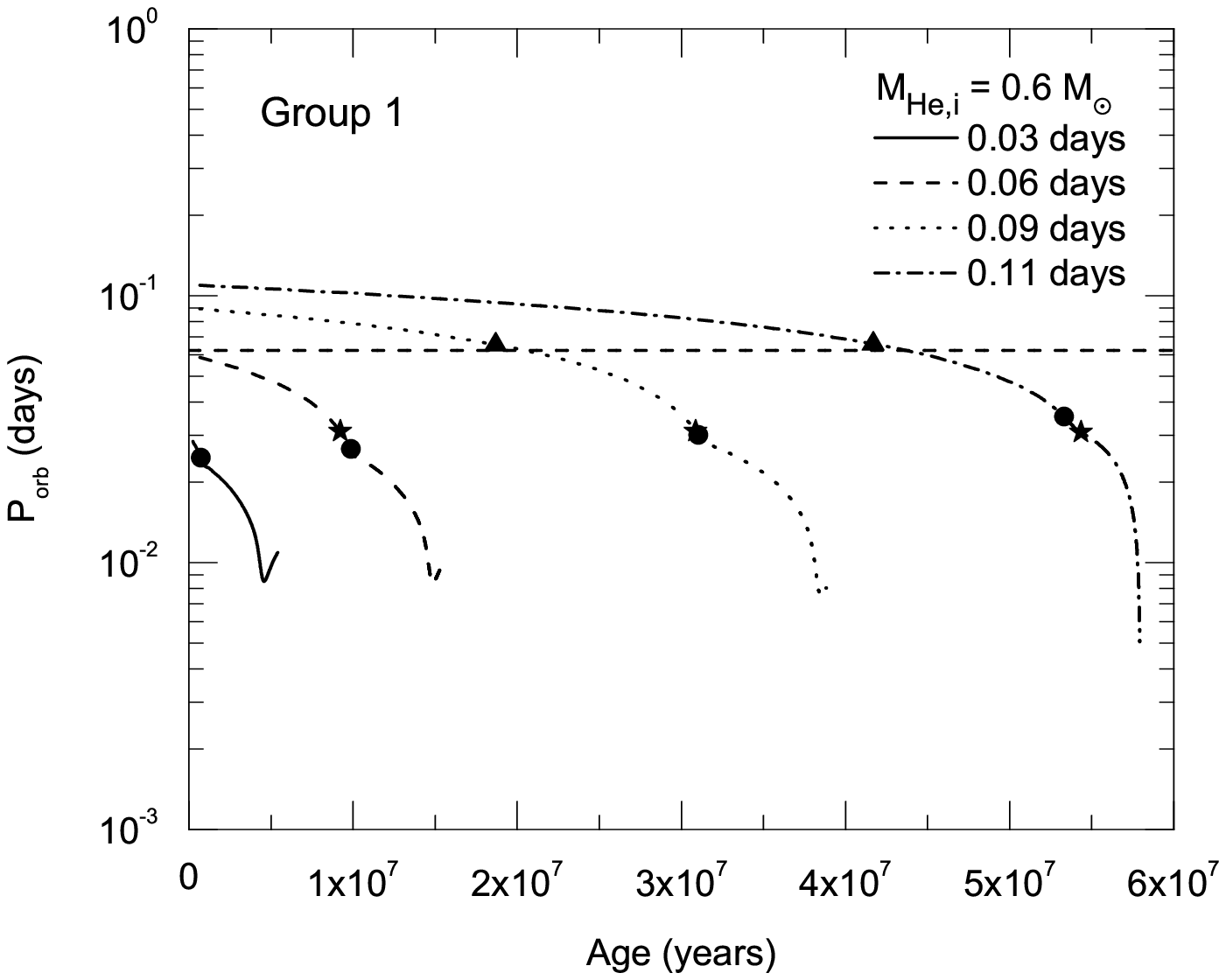}
\includegraphics[width=1.0\linewidth,trim={0 0 0 0},clip]{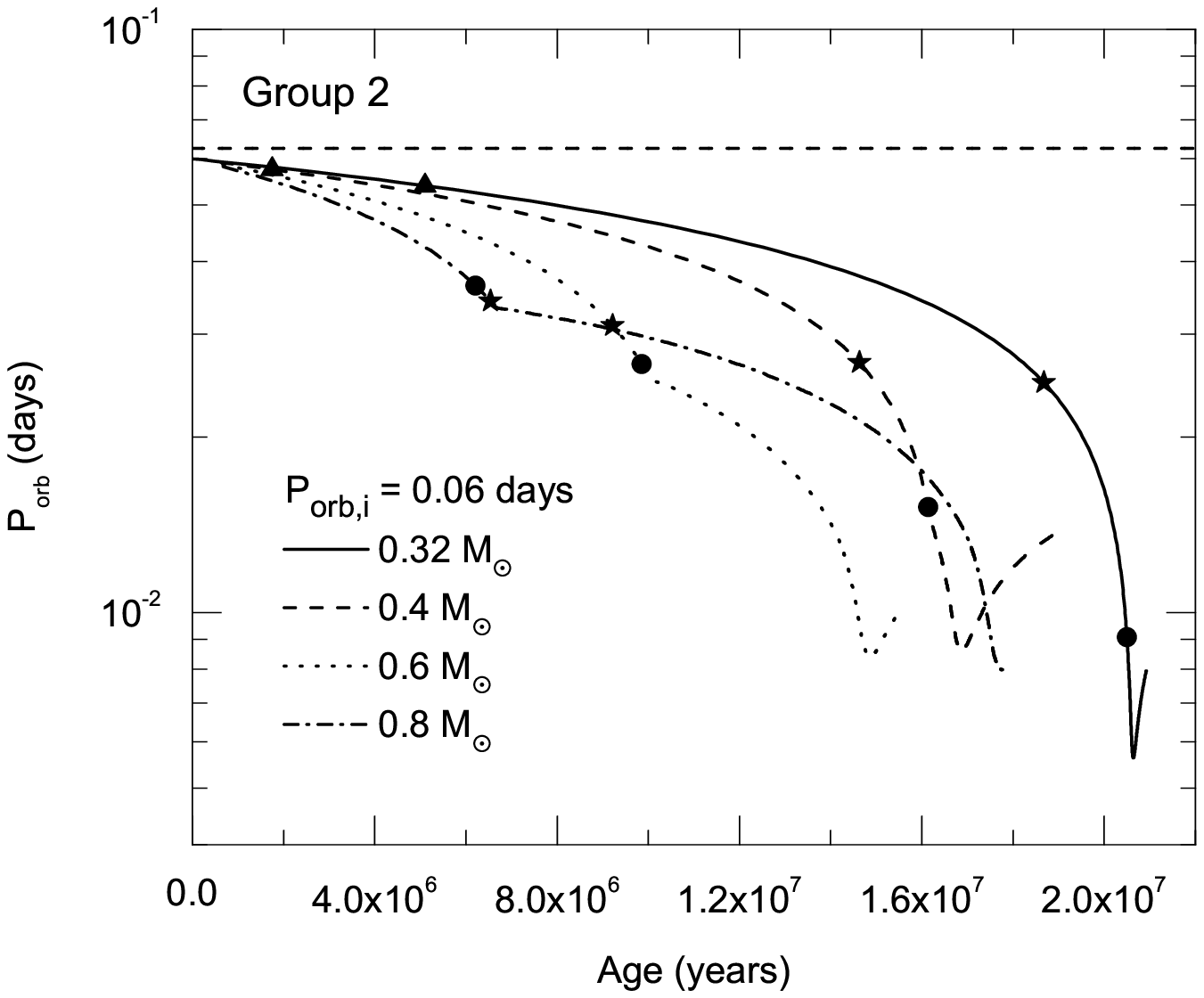}
\includegraphics[width=1.0\linewidth,trim={0 0 0 0},clip]{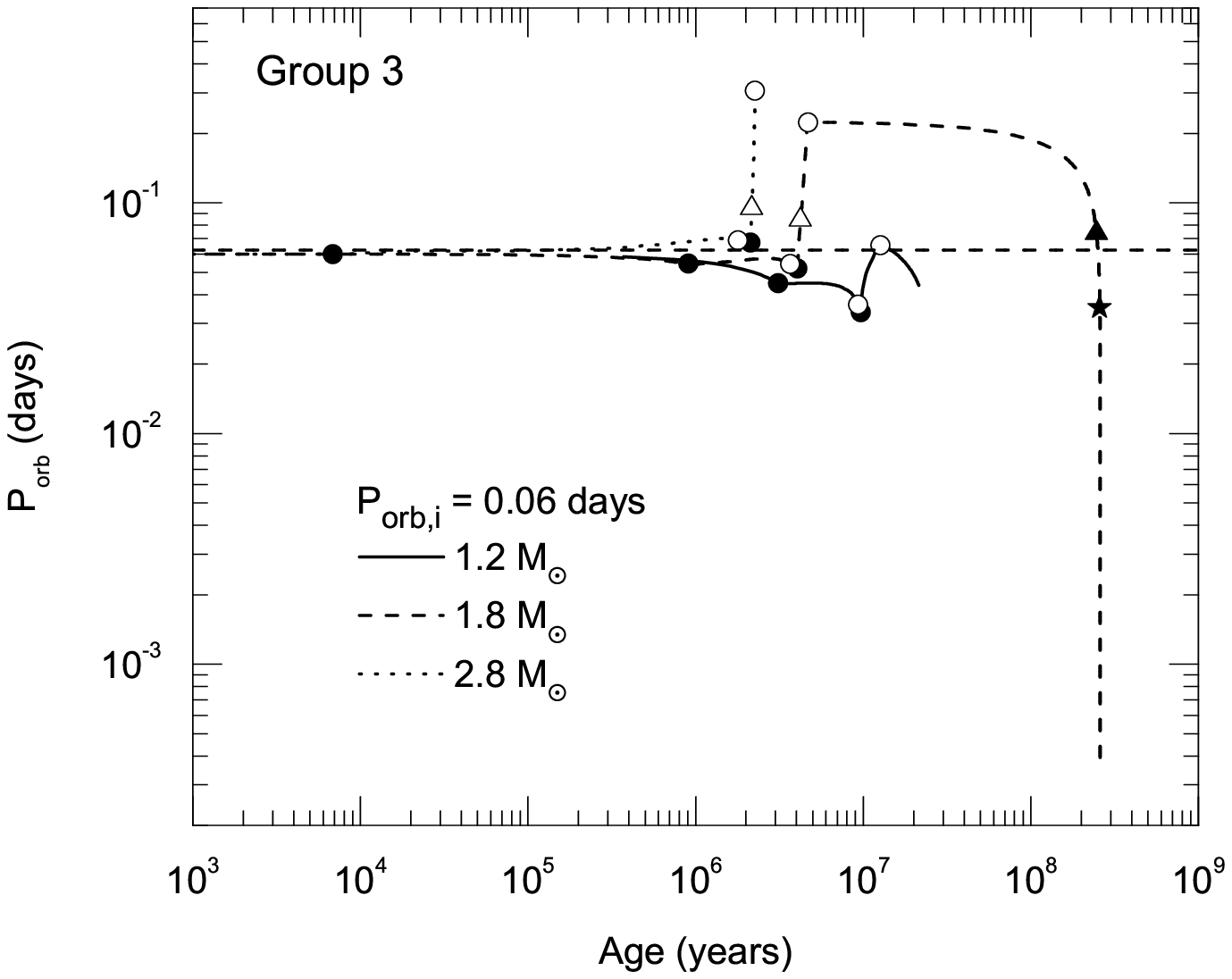}
\caption{Evolution of BH+He star binaries in the orbital period vs. stellar age diagram for groups 1, 2, and 3. The solid circles and open circles denote the onset and end of mass transfer, respectively. The solid triangles and stars represent the onset that BH binaries can be detected by LISA at distances of $d=10$ and 100 kpc,
respectively. BH binaries can not be detected by LISA at a distance of 10 kpc at the open triangles. The horizontal dashed lines represent the threshold period (90 minutes) of UCXBs. } \label{fig:orbmass}
\end{figure}

\begin{figure}
\centering
\includegraphics[width=1.10\linewidth,trim={0 0 0 0},clip]{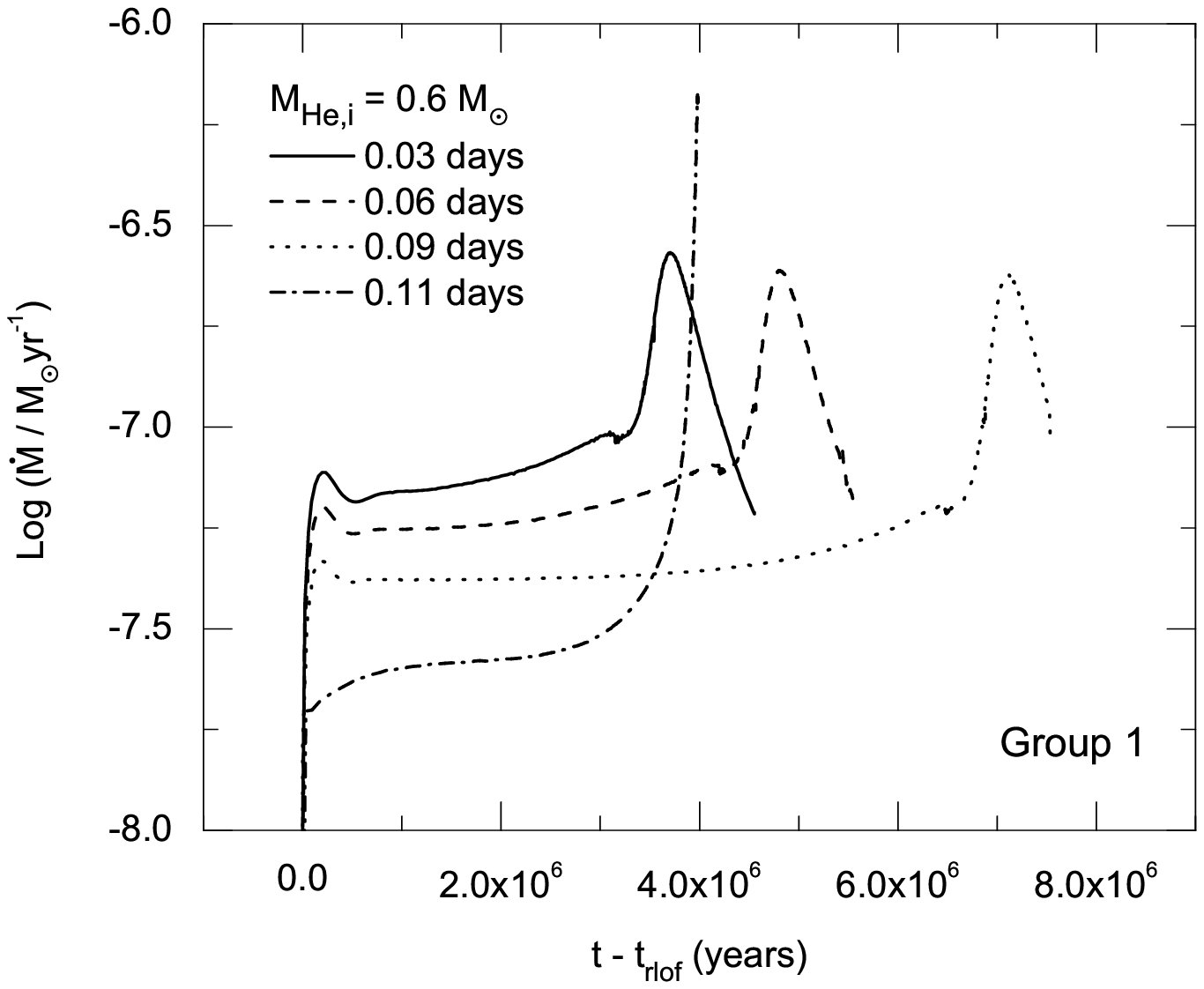}
\includegraphics[width=1.10\linewidth,trim={0 0 0 0},clip]{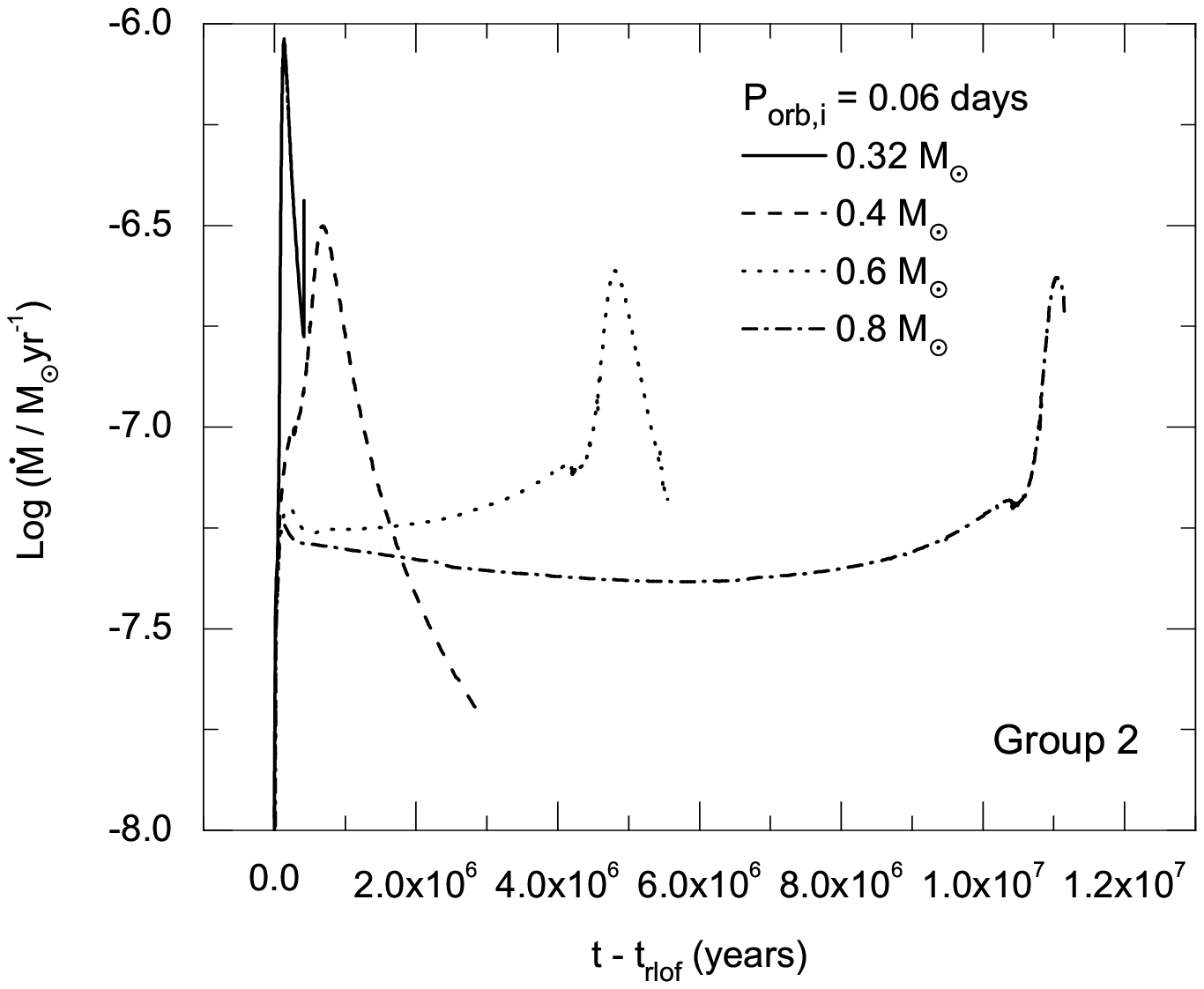}
\includegraphics[width=1.10\linewidth,trim={0 0 0 0},clip]{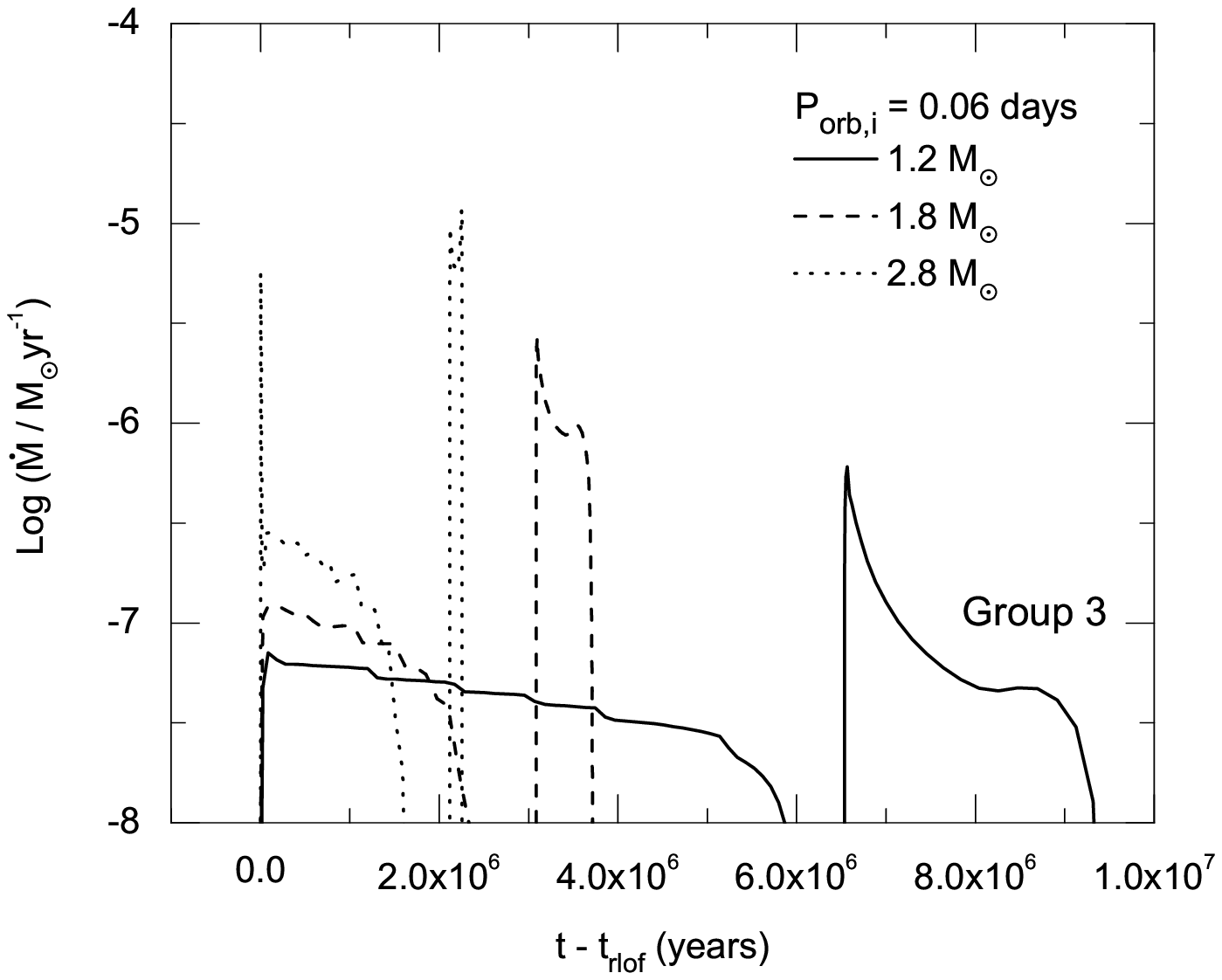}
\caption{Evolution of mass-transfer rate of BH binaries in groups 1, 2, and 3 in the mass-transfer rate vs. mass-transfer timescale diagram. $t$, and $t_{\rm rlof}$ represent the stellar age and the time when the donor star begins RLOF, respectively.} \label{fig:orbmass}
\end{figure}

\subsection{Evolution of mass-transfer rates}
Figure 2 plots the evolution of the mass-transfer rates of BH X-ray binaries in three groups. At $t_{\rm rlof}\approx0.007-54.9~\rm Myr$ (see also Table 1), the He stars fill their Roche lobes. The mass transfer begins early for those massive He stars. Because of a positive correlation between the mass and the radius \citep[$R_{\rm He}\propto M_{\rm He}^{0.654}$, see also][]{taur06} for zero-age MS He stars, the more massive He star is easy to fill its Roche lobe.

For low-mass He stars with $M_{\rm He,i}\leq 0.8~M_{\odot}$, GW radiation dominates the orbital evolution in the early mass-transfer stage. Due to the shrinkage of the orbit, the GW-radiation timescale decreases, thus the mass-transfer rate slowly enhances. After the mass transfer dominates the orbital evolution, the orbital period achieves a minimum, at which the mass-transfer rate emerges a maximum \citep[$10^{-7}-10^{-6}~ M_{\odot}\, \rm yr^{-1}$,][]{yung08}. Subsequently, the He star evolves to more degenerate, and the correlation between the mass and the radius is negative, resulting in a decreasing mass-transfer rate \citep{avil93,yung08}. For high-mass He stars with $M_{\rm He,i}\geq 1.2~M_{\odot}$, mass transfer dominates the orbital evolution due to high mass-transfer rates. The orbital periods increase or slowly decrease, producing decreasing mass-transfer rates.

Three BH+He star binaries in group 3 experience mass transfer twice. The first mass transfer last a relatively long timescale ($1.6-5.8~\rm Myr$) at a rate of $\sim 10^{-7}~\dot{M}_\odot \rm \,yr^{-1}$, which increases with the increase of $M_{\rm He,i}$. As the He abundance in the core drops below 0.1, it develops a carbon-oxygen (CO) core. Due to the contraction of the He star, the binary gradually becomes a detached system, and the first mass transfer ceases. After the core He is exhausted, the He star begins the He-shell burning. With the continuous growth of CO core, the He star begins to expand and initiates the second mass transfer. The mass-transfer rate in the second stage is significantly higher than that of the first mass-transfer stage, and the duration is shorter. A high donor-star mass tends to produce a high mass-transfer rate and a short mass-transfer duration. For $M_{\rm He,i}$ = 1.8 and 2.8 $M_{\odot}$, the second mass-transfer rate can exceed $10^{-6}~\dot{M}_\odot \rm \,yr^{-1}$. The orbital-expand effect caused by such a high $\dot{M}_{\rm tr}$ conquers the orbital-shrinkage effect caused by GW radiation, and the binary orbits begin to widen (see also Figure 1), then the mass-transfer rate continuously decreases. The whole burning He-shell is almost stripped in the second mass-transfer stage, and the remaining CO core becomes highly degenerate. Because of a negative correlation between the mass and the radius of a degenerate star, the binary detaches again. In fact, the BH+He star systems in group 3 may undergo a third RLOF, in which the CO WD fills the Roche lobe and triggers a mass transfer. Due to the limitation of the minimum time step, the duration of the third mass transfer is too short ($\le$ 1 yr) to show in Figure 2.

It is noteworthy that the BH+He star system with $M_{\rm He,i} = 2.8~M_{\odot}$ emerges a transient ultra-high mass-transfer rate ($\sim10^{-5}~\dot{M}_\odot \rm \,yr^{-1}$) in the first mass-transfer stage. The mass transfer proceeds in a thermal time-scale \citep{ergm90,yung08}, thus the mass-transfer rate is very high in this stage, which is much higher than $\dot{M}_{\rm Edd}$. With a rapid decrease of the convective-core mass, the He star rapidly contracts until it re-reaches thermodynamic equilibrium \citep{dewi02}. Subsequently, the mass transfer will continue steadily on the nuclear timescale.

\subsection{Evolution of X-ray luminosities}
Compared with detached binaries, BH UCXBs are ideal multimessenger sources for the detection in both GW and electromagnetic wave bands. X-ray observations of these GW sources could reveal more information about the history of binary evolution \citep{madh08,patr08}, and provide some constraints on the nature of the companion, the structure of accretion disk, the relatively accurate position \citep{liu13,motc14,yao19}, and so on. In the mass transfer phase, the X-ray luminosity of the accretion disk surrounding a BH can be calculated by \citep{kodi02}
\begin{equation}
L_{X}=\left\{
\begin{array}{cl}
\epsilon\dot{M}_{\rm acc}c^{2},\quad \dot{M}_{\rm crit}<\dot{M}_{\rm acc}\leq\dot{M}_{\rm Edd}\\
\epsilon\left(\frac{\dot{M}_{\rm acc}}{\dot{M}_{\rm crit}}\right)\dot{M}_{\rm acc}c^{2},\quad \dot{M}_{\rm acc}<\dot{M}_{\rm crit}\\
\end{array}\right.
\end{equation}
where $\epsilon$ is the radiation efficiency of the accretion disk, $\dot{M}_{\rm crit}$ is the critical accretion rate depending on the transition between the low/hard and high/soft states. In the calculation, we take $\epsilon=0.1$, and $\dot{M}_{\rm crit}=10^{-9}~ M_{\odot}\,\rm yr^{-1}$ \citep{nara95}.

\begin{figure}
\centering
\includegraphics[width=1.15\linewidth,trim={0 0 0 0},clip]{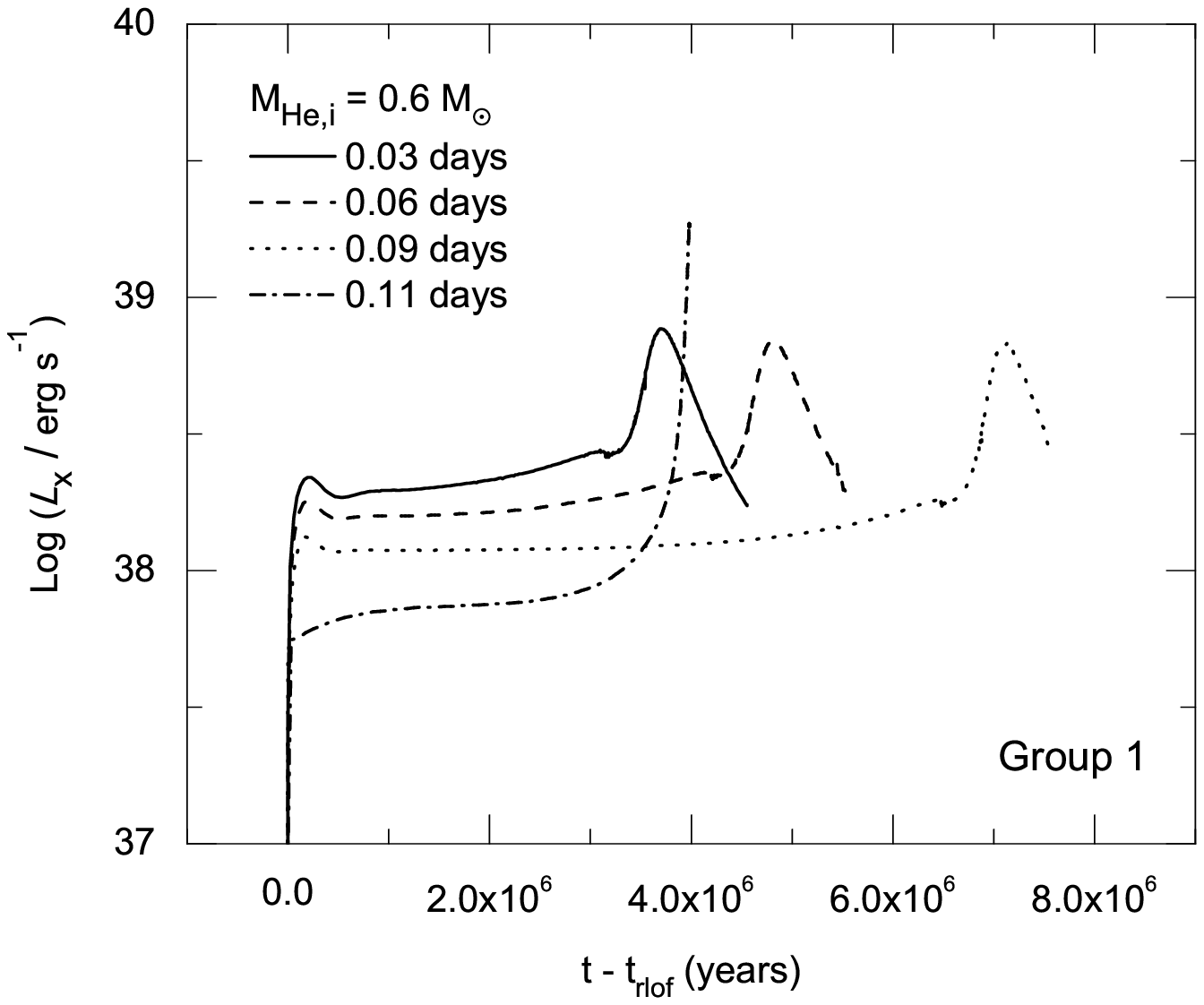}
\includegraphics[width=1.15\linewidth,trim={0 0 0 0},clip]{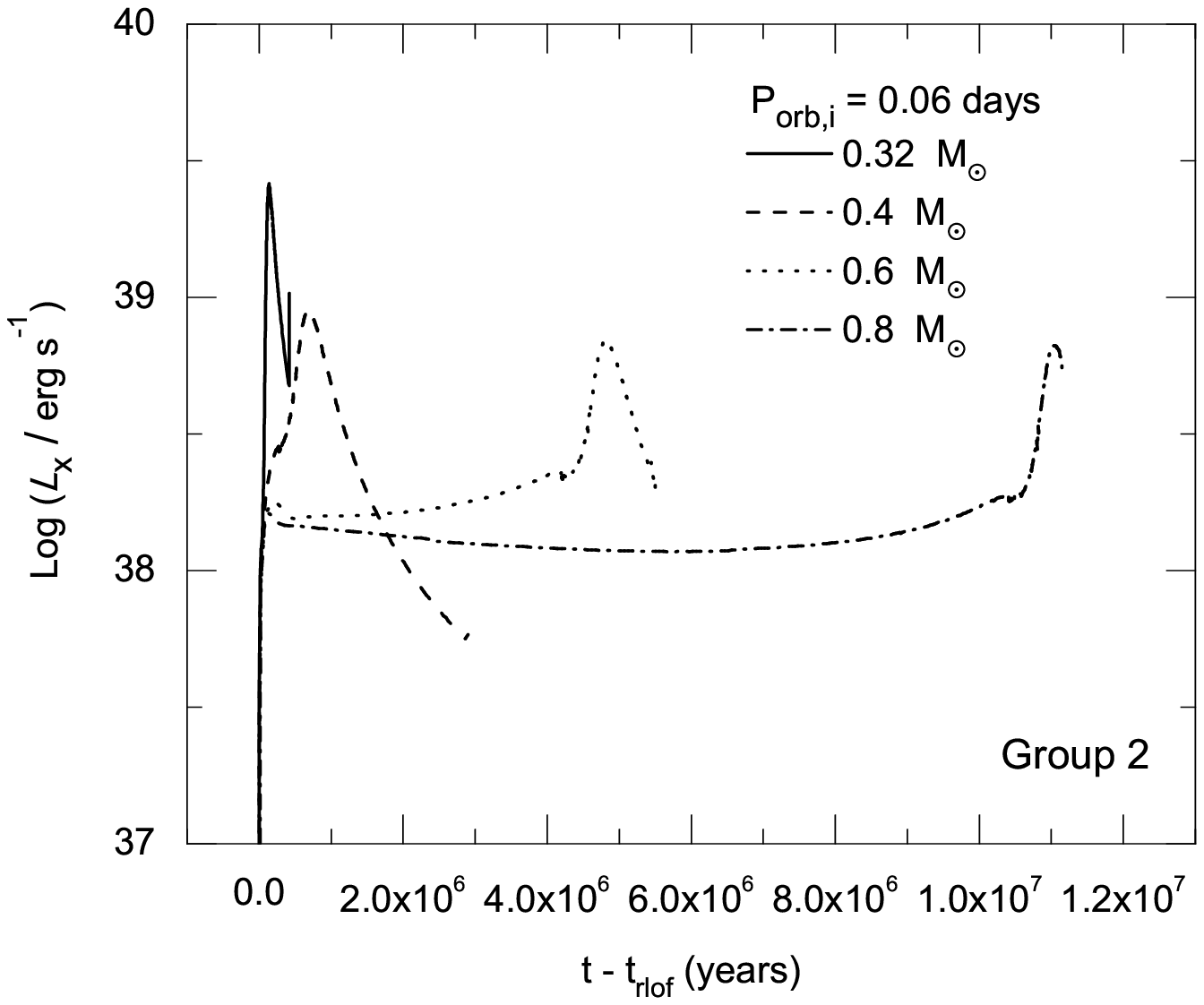}
\includegraphics[width=1.15\linewidth,trim={0 0 0 0},clip]{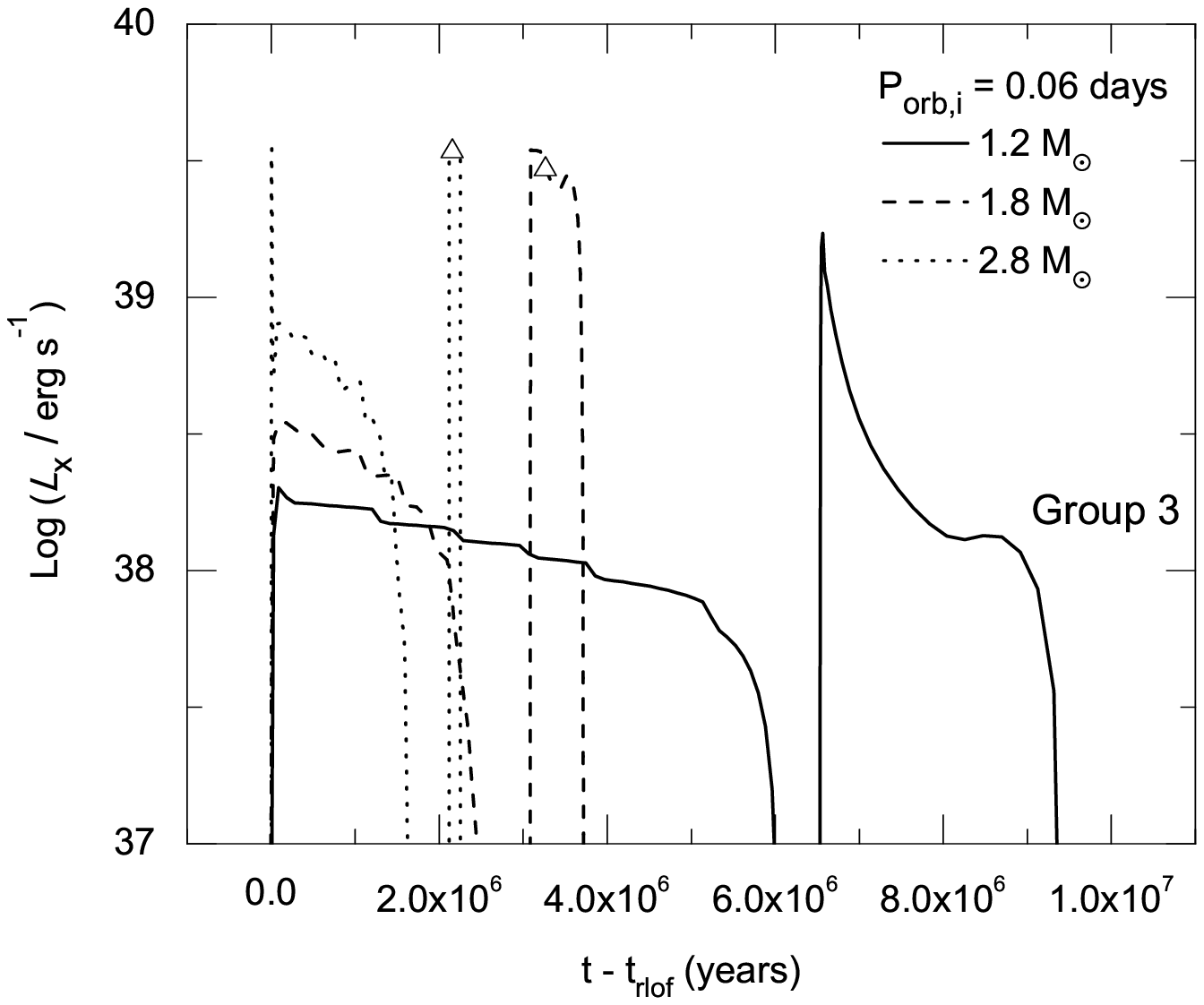}
\caption{Evolution of X-ray luminosities of BH binaries in groups 1, 2, and 3. $t$, and $t_{\rm rlof}$ represent the stellar age and the time when the donor star fills its Roche lobe, respectively. BH UCXBs will not be detected by LISA at a distance of 10 kpc
after the open triangles.} \label{fig:orbmass}
\end{figure}

Figure 3 depicts the evolution of X-ray luminosities of BH UCXBs. It is noteworthy that BH UCXBs evolved from the He star channel produce relatively high X-ray luminosities of $\sim10^{38-39}~\rm erg\, s^{-1}$, which is $2-5$ orders of magnitude higher than those in the MS channel \citep[$\sim10^{33-36}~\rm erg\, s^{-1}$, see also][]{qin23}. However, the timescales of BH UCXBs evolved from the He star channel are $\sim1-10~\rm Myr$, which is much shorter than those ($\sim50-900~\rm Myr$) in the MS channel. All systems in groups 1 and 2 are visible LISA sources at a distance of 10 kpc or 100 kpc during the UCXB stage. Two systems with high donor-star masses ($M_{\rm He,i}$ = 1.8 and 2.8 $M_{\odot}$) in group 3 can not be detected by LISA in the later stage of the second mass-transfer phase because of a rapid orbital expansion. When the binaries evolve to the minimum orbital period, the mass-transfer rates reach the maxima, and the X-ray luminosities emerge peaks ($\sim10^{39}~\rm erg\, s^{-1}$). For a high $M_{\rm He,i}$, BH UCXBs tend to become so-called ultraluminous X-ray sources (ULXs, $L_{X}\geq 10^{39}~\rm erg\, s^{-1}$) \citep{feng11,kaar17}, in which the maximum X-ray luminosity is $3.5\times 10^{39}~\rm erg\, s^{-1}$. Recently, \cite{zhou23} found the first smoking gun evidence for the existence of He donor star in ULXs by Very Large Telescope Multi Unit Spectroscopic Explorer observations. Similarly, the population synthesis study also indicated that NS X-ray binaries containing the He stars can account for a large part of ULXs in Milky Way-like galaxies \citep{shao19}.

\subsection{Detection of GW signals}
Since the Laser Interferometer Gravitational-Wave Observatory (LIGO) first detected the high-frequency GW signal from the double BH merger event GW150914 \citep{abbo16}, the detection of GW opens a new window to understand the distant universe. GW signals provide us with more useful information about stellar and binary evolution. Figure 4 plots the evolution of BH+He star systems in the characteristic stain versus GW frequency diagram. Because of the same chirp mass, the evolutionary tracks of the four systems in group 1 overlap before RLOF. Due to short initial orbital periods, the nascent BH+He star system with $P_{\rm orb,i}=0.03,$ and 0.06 days can be instantly detected by LISA and Taiji at a distance of 10 kpc after the CE stage. It is impossible to form detached BH+WD systems via isolated binary evolution \citep{qin23}. Therefore, those sources with chirp masses similar to those of BH+He star systems should be the post-CE systems, and provide indirect evidence of CE evolutionary stages. When the mass transfer starts, these low-frequency GW sources also appear as UCXBs, which are the ideal multimessenger detection sources. For a long detection distance of 100 kpc, those binaries have to spend a longer evolutionary timescale to evolve toward a shorter orbital period, thus appearing as GW sources only in a few Myr. BH X-ray binaries in Group 3 experience a rapid orbital expansion, thus two systems with $M_{\rm He,i}=1.8$, and $2.8~M_{\odot}$ are not potential GW sources in the later stage of the second mass transfer. However, the GW radiation will drive the orbit of the system with $M_{\rm He,i}=1.8~M_{\odot}$ to continuously shrink after the mass transfer ceases, resulting in a low-frequency GW source that is detectable in a long timescale. Since the chirp masses are constant, the evolutionary tracks of detached systems are lines with a slope of $7/6$ ($h_{\rm c}\propto f_{\rm gw}^{7/6}$ according to equation 3). In groups 1 and 2, the maximum GW frequency of BH UCXBs is 5.6 mHz, while the maximum frequency can reach about 63.5 mHz for a detached BH binary in group 3 (see also Table 1).

\begin{figure}
\centering
\includegraphics[width=1.15\linewidth,trim={0 0 0 0},clip]{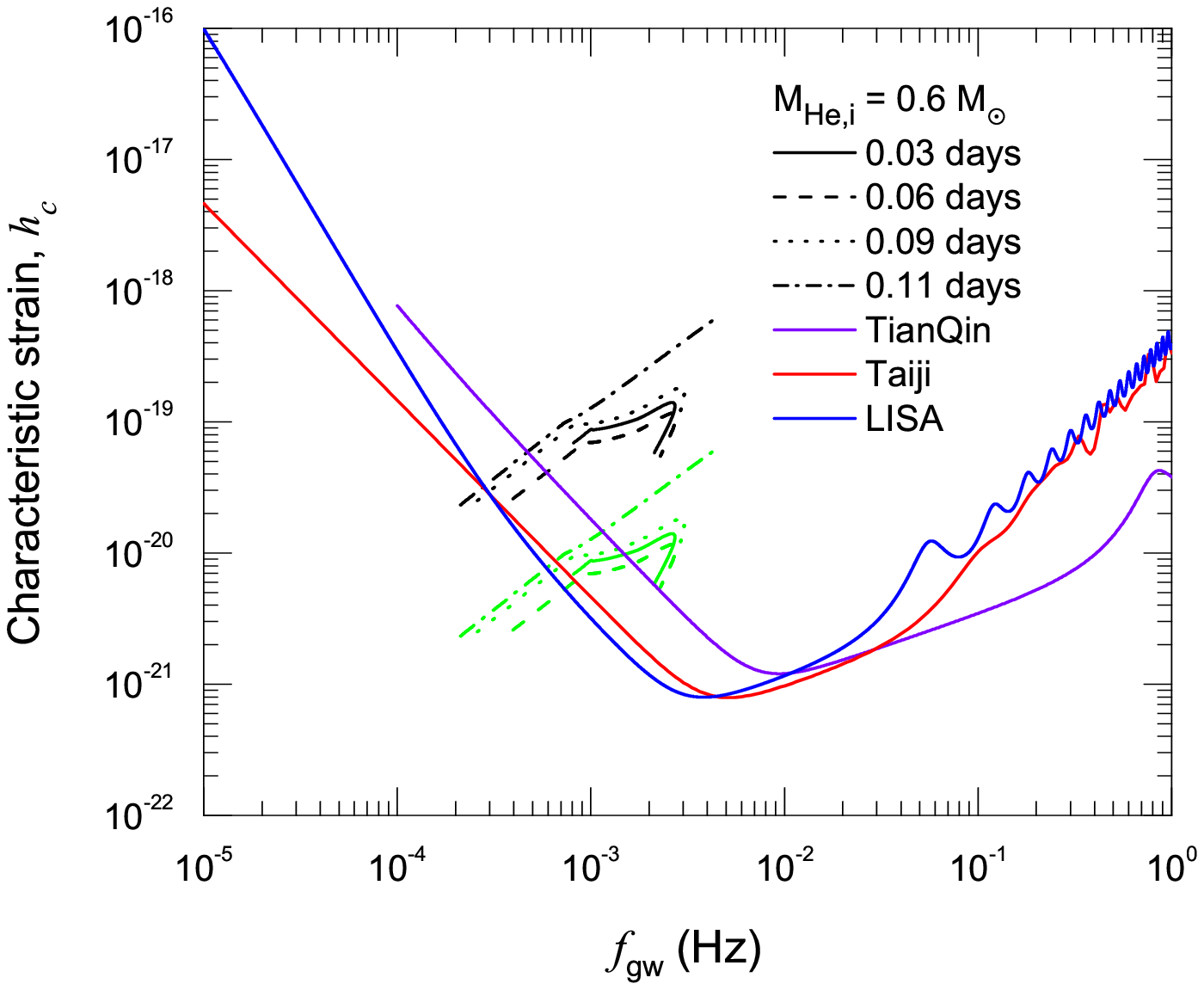}
\includegraphics[width=1.15\linewidth,trim={0 0 0 0},clip]{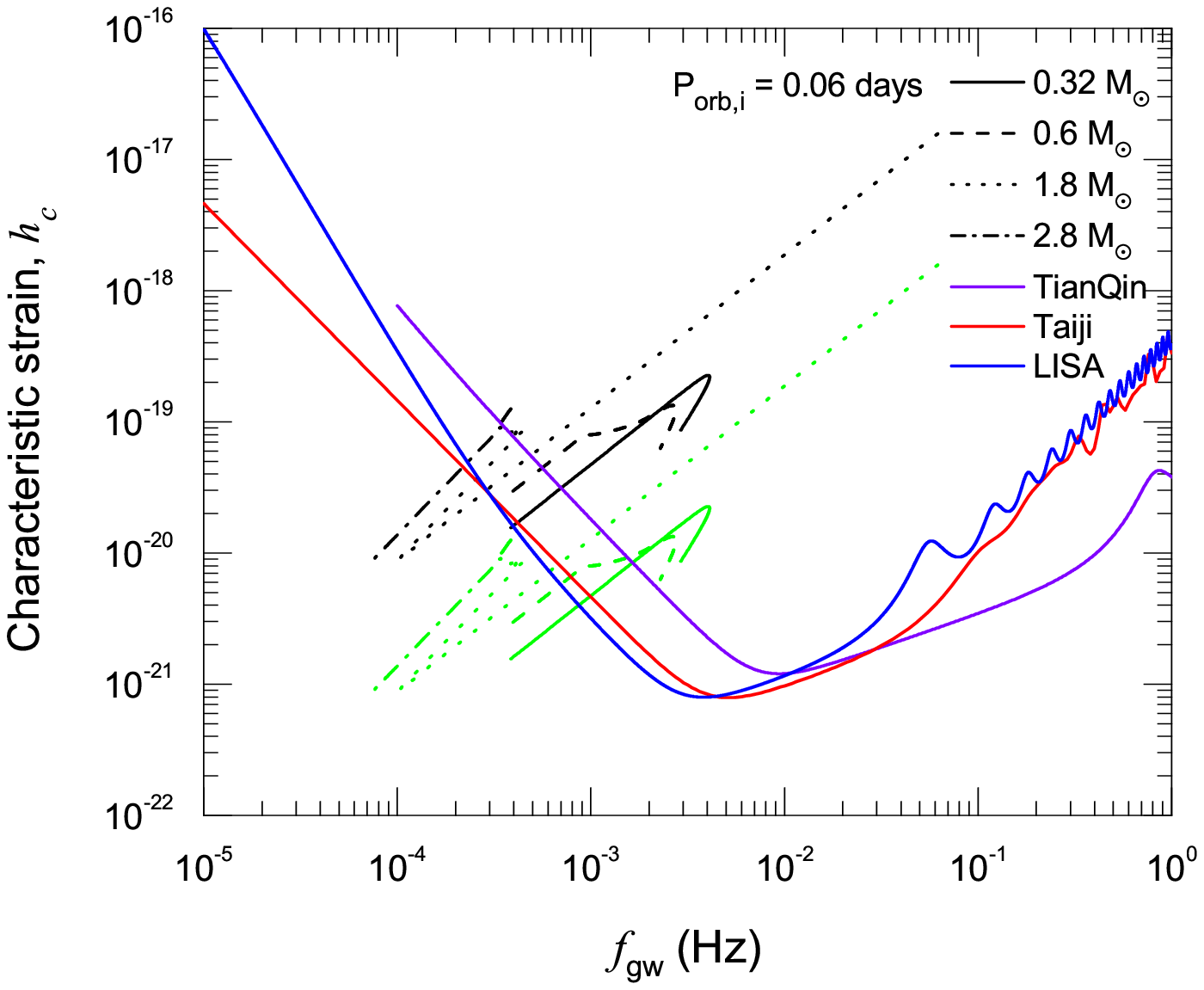}
\caption{Evolution of BH+He star binaries in groups 1, 2, and 3 in the characteristic strain vs. GW frequency diagram. The detection distances of the black and green curve groups are 10, and 100 kpc, respectively. To avoid overlap, the dashed, dotted, and dashed-dotted curves are slightly moved up and down in parallel in the top panel (actually, these three curves overlap with the solid curve). The blue, purple, and red curves denote the sensitivity curve of LISA \citep{robs19}, TianQin \citep{wang19}, and Taiji \citep{ruan20} based on the numerical calculation of 4-year observations, respectively.} \label{fig:orbmass}
\end{figure}

To understand the progenitor properties of BH UCXBs, we model the evolution of a large number of BH+He star binaries. Figure 5 summarizes the initial contour of the progenitors of BH UCXBs in the $P_{\rm orb,i}-M_{\rm He,i}$ diagram. All BH+He star binaries with initial parameters located between two solid curves can evolve toward UCXBs (with durations of $\geq$ 0.1 Myr) that can be detected by LISA at a distance of 10 kpc. The initial He-star masses and initial orbital periods of progenitors of Galactic BH UCXB-LISA sources are in the range of $0.32-2.9~M_{\odot}$ and $0.02-0.19$ days, respectively. Such an initial parameter space is slightly wider than that ($0.32-1.2~M_{\odot}$ and $0.01-0.1$ days) for the progenitors of NS UCXB-LISA sources evolving from the He star channel \citep{wang21}. It is clear that the systems with relatively long initial orbital periods ($P_{\rm orb,i} \geq 0.05~\rm days$) and high donor-star masses ($M_{\rm He,i} \geq 0.7~M_{\odot}$) may experience two RLOF stages. For massive He stars with $M_{\rm He,i} \geq 1.2 M_{\odot}$, all BH binaries will experience two RLOFs. Some systems marked by crosses can also evolve into low-frequency GW sources, however, they can not become valid UCXBs with a duration of $\geq$ 0.1 Myr. Those BH+He star binaries that experienced two RLOF stages could produce a high GW frequency in the final evolutionary stage (see also $M_{\rm He,i}=1.8~M_{\odot}$ in Figure 4), while these sources are not UCXBs because of the absence of mass transfer. For a detection distance of 10 kpc, these binaries are most likely as multimessenger sources in the first mass-transfer stage.

\begin{table*}
\begin{center}
\caption{Some Important Evolutionary Parameters of BH+He Star Binaries in Groups 1, 2, and 3. \label{tbl-2}}
\begin{tabular}{@{}lllllllllll@{}}
\hline\hline\noalign{\smallskip}
$M_{\rm He,i}$ & $P_{\rm orb,i}$ &  $t_{\rm rlof}$ &$P_{\rm orb,min}$ &$t_{\rm orb,min}$ &$f_{\rm gw,max}$ & $L_{\rm X\_10,max}$ &$L_{\rm X\_100,max}$& $\bigtriangleup t_{\rm LISA,10}$& $\bigtriangleup t_{\rm LISA,100}$\\
 ($ M_{\odot}$)     &  (days)  & (Myr)   &   (min) & (Myr)    & (mHz)& ($10^{38}~\rm erg\,s^{-1}$) & ($10^{38}~\rm erg\,s^{-1}$)& (Myr)& (Myr)\\
\hline\noalign{\smallskip}
0.6 & 0.03 &0.852 &12.3&4.58& 2.72& 7.67 & 7.67& 5.42 & 5.42\\
0.6 & 0.06 &10.0  &12.1& 14.8&2.75&6.97 &6.97 & 15.4 & 6.17\\
0.6 & 0.09 &31.4  &11.1&38.9& 3.0&6.48 &6.48&20.3 & 8.13\\
0.6 & 0.11 &54.9  &6.01&58.1&5.55&34.7 &34.7&15.8 &3.43\\
0.32& 0.06 &20.5  &8.12&20.6&4.11& 25.6& 25.6& 15.9 &2.26\\
0.4 & 0.06 &16.2  &12.4&16.9&2.69&8.97 &8.97&17.4 &4.47\\
0.8 & 0.06 &6.69  &11.5&17.7&2.9&6.66 &6.66&17.7 &11.1\\
1.2 & 0.06 &3.13  &48.3&9.67&0.691&33.9 &$-$&21.5 &$-$\\
1.8 & 0.06 &0.962 &0.525&260&63.5&34.6 &$-$&17.5 &$1.82^{*}$\\
2.8 & 0.06 &0.007 &86.3&0.007&0.386&35.1 &$-$&2.16 &$-$\\
\hline\noalign{\smallskip}
\end{tabular}
\tablenotetext{}{}{Note. The columns list (in order): the initial He-star mass, the initial orbital period, the stellar age at the beginning of RLOF, the minimum orbital period, the stellar age at the minimum orbital period, the maximum frequency of GW, the maximum X-ray luminosity when binaries are LISA sources at a distance of 10 kpc and 100 kpc, the detection timescale that the system can be detected by LISA at a distance of 10 and 100 kpc.\\
$*$ The system can be detected by LISA at a distance of 100 kpc as a detached system.}
\end{center}
\end{table*}

\begin{figure}
\centering
\includegraphics[width=1.15\linewidth,trim={0 0 0 0},clip]{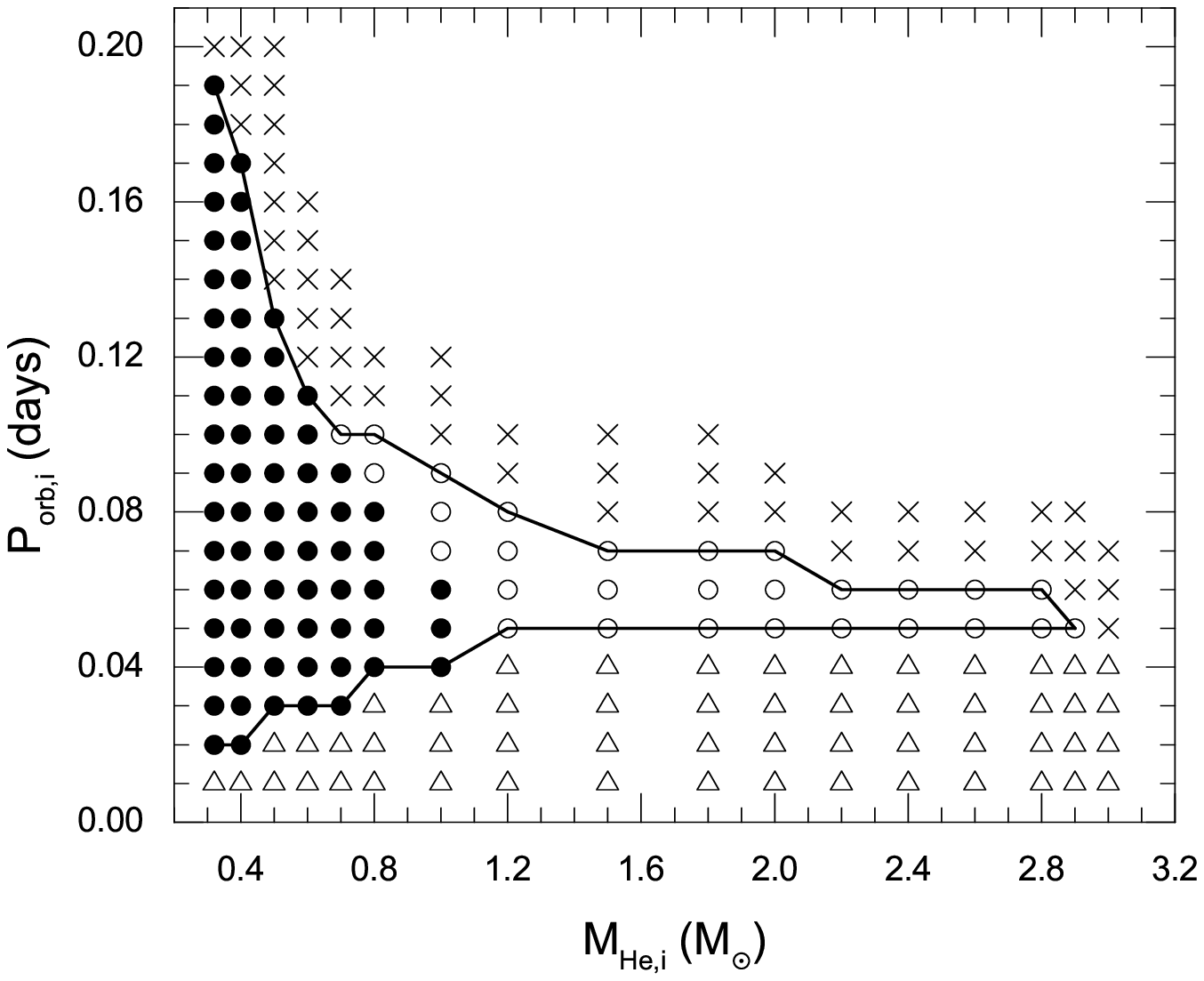}
\caption{Parameter space distribution of BH+He star binaries with different evolutionary fates in the initial orbital period vs. initial He-star mass diagram. The BHs are taken to be a constant mass of 8 $M_{\odot}$. The solid curves represent the boundaries that can evolve into UCXBs (with durations $\geq$ 0.1 Myr) that can be detected by LISA at a distance of 10 kpc. The solid and open circles denote the progenitors of BH UCXB-LISA sources that experience one RLOF, and two RLOF, respectively. The crosses stand for BH+He star systems that cannot evolve into valid UCXBs (with durations $\le$ 0.1 Myr). The open triangles correspond to BH+He star systems that He stars have already filled their Roche lobes at the beginning of evolution.} \label{fig:orbmass}
\end{figure}

The measurement of chirp mass is very significant in constraining the masses of two components. For a detached binary system, the chirp mass can be derived by
\begin{equation}
\mathcal{M}=\frac{c^{3}}{G}\left(\frac{5\pi^{-8/3}}{96}f_{\rm gw}^{-11/3}\dot{f}_{\rm gw}\right)^{3/5},
\end{equation}
where $\dot{f}_{\rm gw}$ is the GW-frequency derivative. Therefore, the measurement of chirp mass depends on the accuracy of the $\dot{f}_{\rm gw}$. The space GW detectors are only able to detect $\dot{f}_{\rm gw}$  for those ultra-compact binaries
with a large signal-to-noise ratio (SNR) and a small orbital period close to the minimum orbital period \citep{taur18}. The minimum $\dot{f}$ that LISA can measure is given by \citep{taka02,taur18}:
\begin{equation}
\dot{f}_{\rm gw,min}\approx 2.5\times 10^{-17}\left(\frac{10}{S/N}\right)\left(\frac{4~\rm yr}{T}\right)^{2}~\rm Hz\,s^{-1},
\end{equation}
where $S/N$ is the SNR of GW signals, and $T$ is the mission duration of LISA.

Figure 6 illustrates the evolution of $\dot{f}_{\rm gw}$ with the GW frequencies for our simulated three groups. Taking $S/N = 10$ and $T = 4$ yr, we have the detection limitation of GW-frequency derivative as $\dot{f}_{\rm gw,min}=2.5\times 10^{-17}~\rm Hz\,s^{-1}$, which is plotted by the horizontal dashed lines in Figure 6. In the orbital shrinkage stages, the angular momentum loss rates by GW radiation continuously increase, resulting in a rapid increase of $\dot{f}_{\rm gw}$. The maximum $\dot{f}_{\rm gw}$ is close to the peak GW frequency, however, $\dot{f}_{\rm gw}$ sharply decreases to be 0 at the maximum GW frequency, then its sign turns into negative due to an orbital expansion. Except for the orbital expansion stage of two systems in group 3, other BH UCXBs could provide detectable $\dot{f}_{\rm gw}$ in a timescale of $1~\rm Myr$. All evolutionary curves of $\dot{f}_{\rm gw}$ in the climbing stage are approximate lines with a slope of $n=11/3$, which implies a relation $\dot{f}_{\rm gw}\propto f_{\rm gw}^{11/3}$. This is consistent with that the GW radiation dominates the orbital evolution of the binaries. According to equation (6), $\dot{f}_{\rm gw}\propto f_{\rm gw}^{11/3}$ for a constant chirp mass in detached binaries, in which the angular-momentum loss is fully contributed by GW radiation \citep{webb98,piro19}. Therefore, if a braking index as $\dot{f}_{\rm gw}\propto f_{\rm gw}^{n}$ is defined, its measurement can diagnose whether the binary emitting low-frequency GW signals is a detached system.

Figure 7 presents the parameter space of BH+He star systems whose chirp masses can be accurately measured. For BH UCXB-LISA sources evolving from the He star channel, only several systems evolved from the progenitors with $M_{\rm He,i}= 0.7-1.5 ~M_{\odot}$ and $2.4-2.6 M_{\odot}$ and relatively long $P_{\rm orb,i}$ are difficult to detect $\dot{f}_{\rm gw}$. For the MS channel, only BH UCXBs with initial orbital periods very near the bifurcation period have detectable $\dot{f}_{\rm gw}$ \citep{qin23}. Therefore, BH UCXB-LISA sources evolved from the He star channel are most likely to measure their chirp masses.

\begin{figure}
\centering
\includegraphics[width=1.15\linewidth,trim={0 0 0 0},clip]{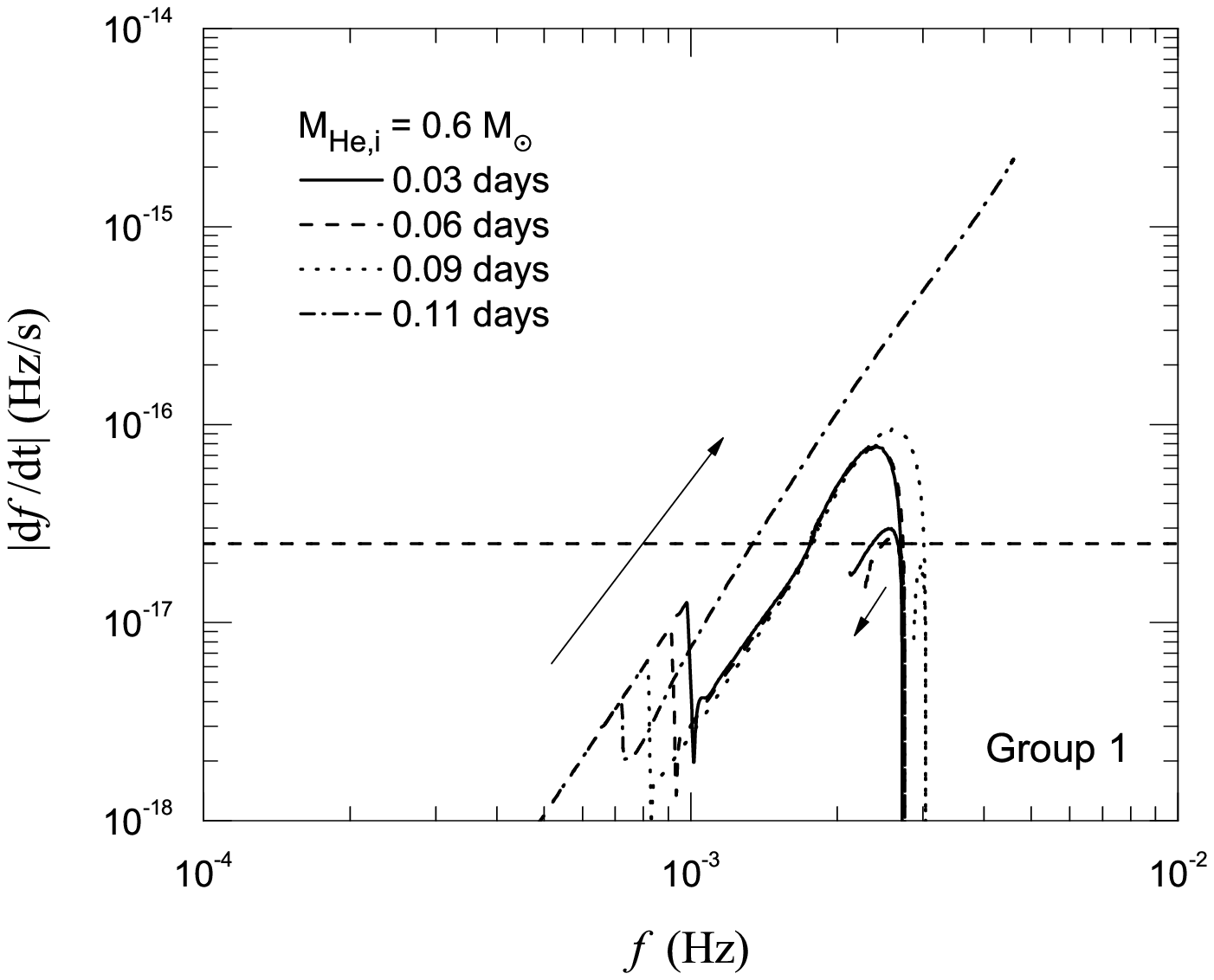}
\includegraphics[width=1.15\linewidth,trim={0 0 0 0},clip]{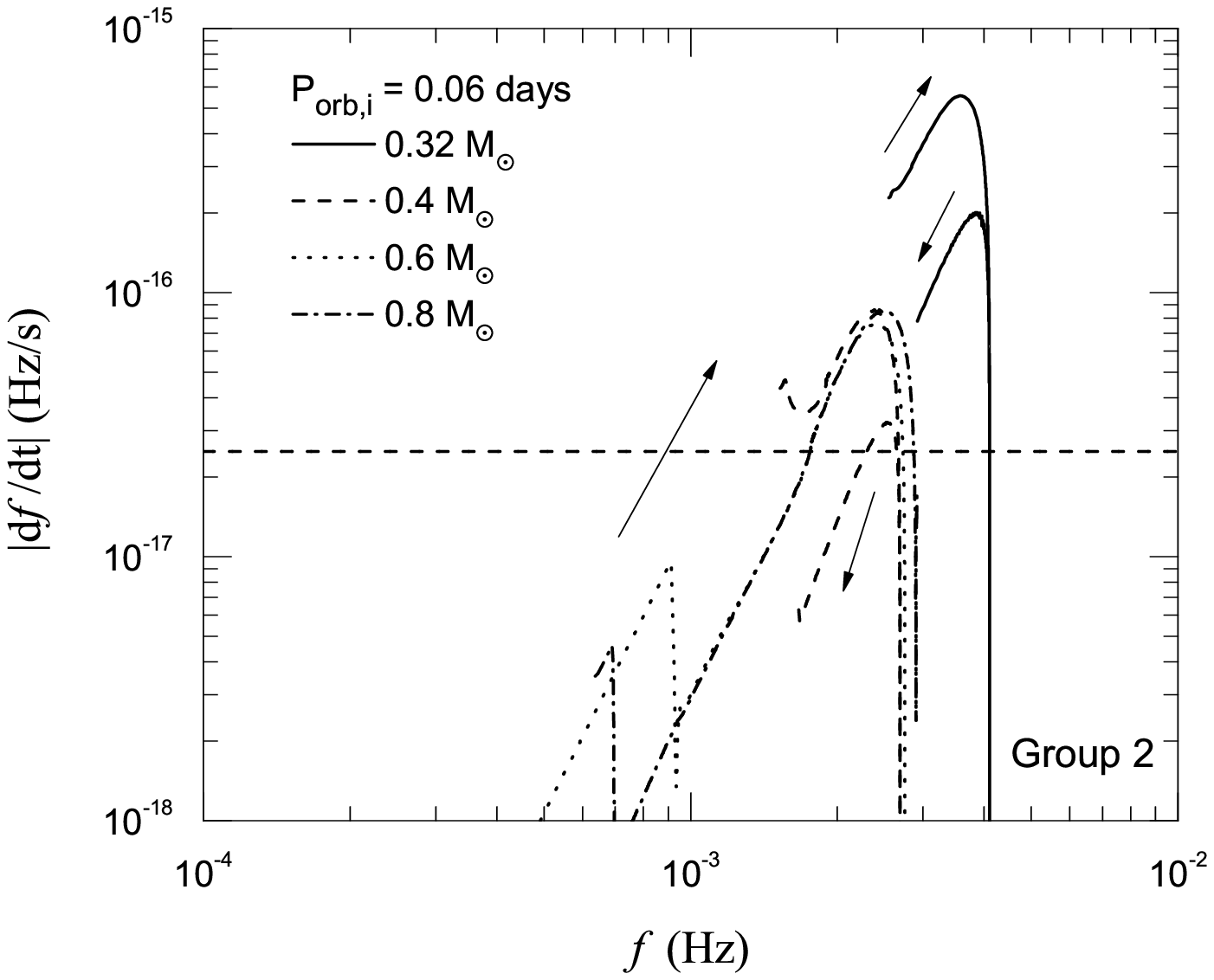}
\includegraphics[width=1.15\linewidth,trim={0 0 0 0},clip]{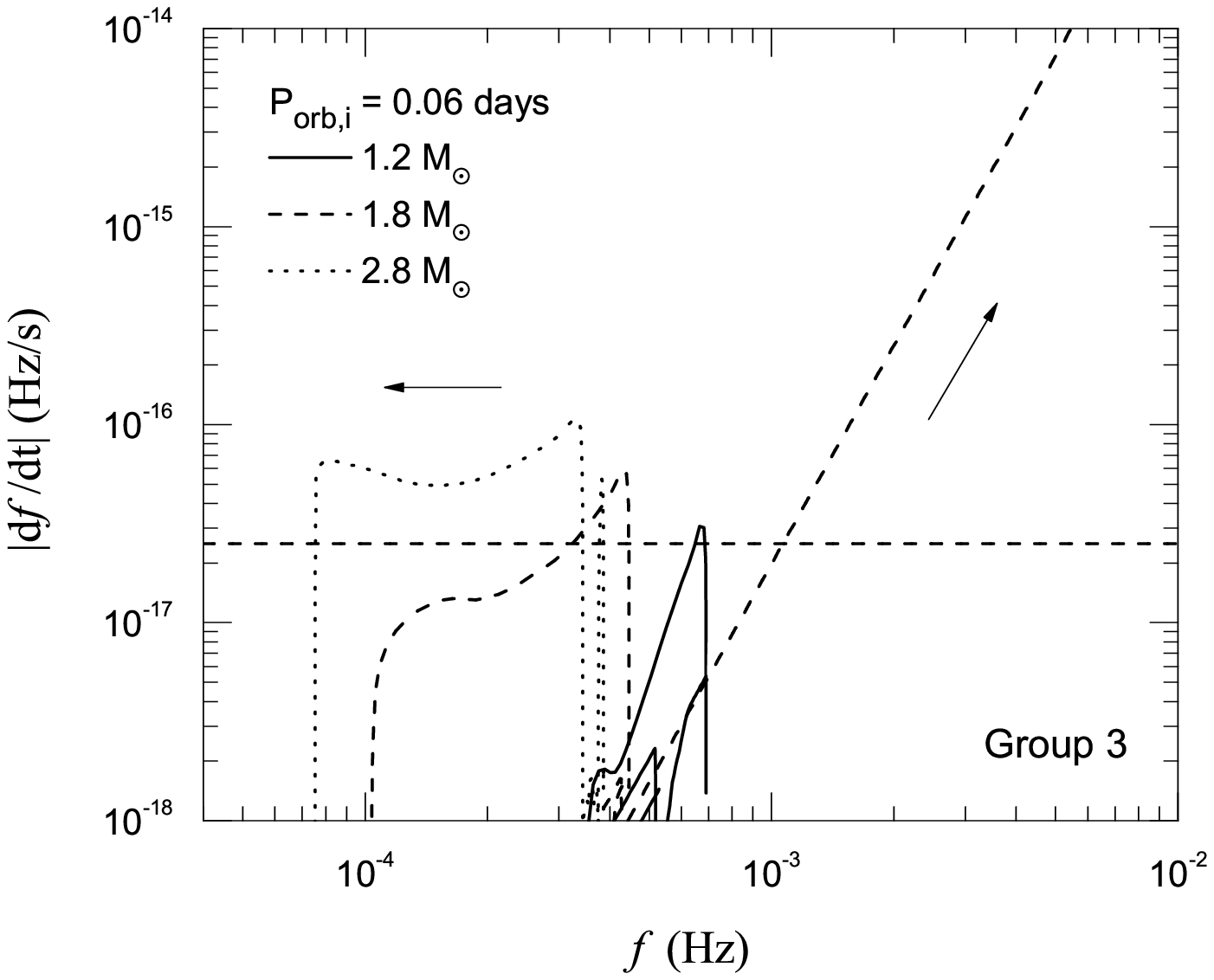}
\caption{Evolution of GW frequency derivative of BH UCXBs in groups 1, 2, and 3 in the $| \dot{f}_{\rm gw}|$ vs. $f_{\rm gw}$ diagram. The horizontal dashed line represents the minimum $\dot{f}_{\rm gw}$ that can be detected by LISA for $S/N = 10$ and $T = 4~\rm yr$ (see also equation 7).} \label{fig:orbmass}
\end{figure}

\begin{figure}
\centering
\includegraphics[width=1.15\linewidth,trim={0 0 0 0},clip]{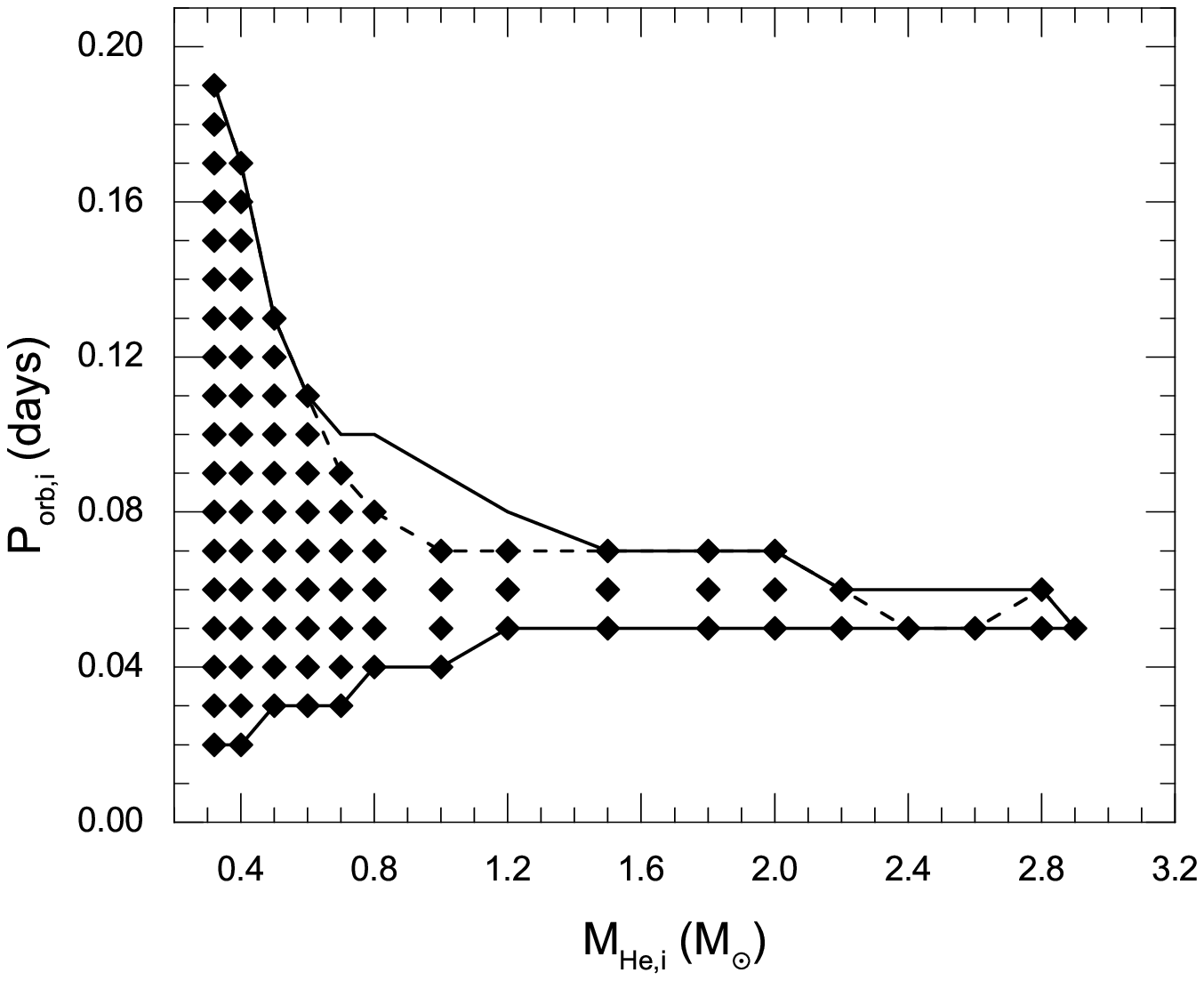}
\caption{Parameter space distribution that describes the detectability of $\dot{f}$ for BH+He star systems in the initial orbital period vs. initial He-star mass diagram. The solid diamonds represent BH+He star systems whose $\dot{f}$ can be detected by LISA during the UCXB stage. The solid curves represent the boundary that can evolve toward the valid UCXB stage (durations $\geq$ 0.1 Myr). The dashed curves represent the boundary that $\dot{f}$ can be detected (the lower boundary coincides with the solid curve).} \label{fig:orbmass}
\end{figure}

\section{Discussion}
\subsection{Origin of BH+He star binaries}
Similar to NS+He star systems, BH+He star systems could also be descendants of high-mass X-ray binaries, in which the companion of the BH loses its hydrogen envelope through stellar wind or a mass-transfer stage \citep{bhat91,dewi02}. Cyg X-3 was proposed to include a $2-4.5~M_{\odot}$ BH (or NS) and a $7.5-14.2~M_{\odot}$ Wolf-Rayet donor star \citep{zdzi13}, and it may be the progenitor of a Galactic double BH or BH-NS \citep{belc13}. In addition, massive He stars might be formed through quasi-chemically homogeneous evolution \citep{woos06,yoon06,cant07}. However, our simulations find that BH binaries with massive He stars ($M_{\rm He,i}>3.0~M_{\odot}$) are hard to evolve into detectable UCXBs with a duration longer than 0.1 Myr. \cite{jian23} performed 1D model of post-CE BH binaries with short orbital period ($\leq0.2$ days) consisting of a BH and a massive He star ($M_{\rm He,i}\geq3.3~M_{\odot}$) that experiences stable mass transfer, and found that their mass-transfer timescales are shorter than 0.1 Myr, which is consistent with our results.

NS/BH+He star binaries could evolve from NS/BH+MS binaries through Case B or Case C mass transfer \citep{dewi02}. For those systems with a large mass ratio, the mass transfer is dynamically unstable. Subsequently, the binary systems enter a
CE phase \citep{ivan13}, and the evolutionary products of those systems that experienced Case B mass transfer in the CE phase are binaries with an unevolved He star and a NS/BH. Those NS/BH+He star binaries produced from Case B are so close that they initiate a Case BA mass transfer in the core He burning stage. It is clear that our simulated systems in the parameter space of the BH UCXB-LISA source are formed from Case B mass transfer. Case C mass transfer would produce an evolved He star and a NS/BH after the CE phase, subsequent mass transfer would start after core He is exhausted.

\subsection{He star Channel and MS channel}

There exists a bifurcation period for the formation of UCXBs in the MS channel, which is defined as the maximum initial orbital period forming UCXBs within a Hubble time \citep{sluy05,seng17,chen20,qin23}. However, in the He star channel there is not a critical period similar to the MS channel because of short initial orbital periods. In the He star channel, the initial donor-star masses of the BH+He star system that can evolve into UCXB-LISA sources are in a wide range from 0.32 to $2.9~M_{\odot}$. However, this range is shortened to be $0.4-1.6~M_{\odot}$ in the MS channel because the stars with radiative envelope will not experience magnetic braking \citep{qin23}. Certainly, the initial orbital period range of the He star channel is much narrower than that of the MS channel. In the MS channel, the maximum GW frequency emitting by BH UCXBs is $\sim 3~\rm mHz$ \citep{qin23}, which is slightly smaller than that ($5.6~\rm mHz$) in the He star channel. Because of the high compactness of He stars, BH+He star systems can evolve into LISA sources that can be detected at a distance of 100 kpc, while this phenomenon is impossible for the MS channel.

To understand the final evolutionary fates of the He stars, we plot the evolution of five BH-He star binaries
in an H-R diagram in Figure 8. The final luminosities and the effective temperatures of five He stars are ${\rm log}(L/L_{\odot})\sim-1$ to $-3$, and $6300-31000~\rm K$, respectively. Such luminosity and effective temperature ranges are comparable to those of WDs. It is clear that those BH binaries with a massive donor star ($1.2-1.8~M_{\odot}$) had already evolved into detached systems before the donor stars evolved into WDs by a contraction and cooling stage. Therefore, the He star channel could form detached BH-WD systems. It is noteworthy that the MS channel can only form mass-transferring BH-WD systems rather than detached BH-WD systems \citep{qin23}. Similar to the MS channel, BH binaries with a low-mass ($0.32-0.6~M_{\odot}$) He star can only evolve into semi-detached BH-WD binaries rather than detached systems. Because those WDs evolving from the He star channel lack a thin H envelope, their effective temperatures are much higher than those of WD evolving from the MS channel.

\subsection{He star Channel and Dynamic Process Channel}
In globular clusters and young dense clusters, compact BH-WD binaries might assembled through BHs capturing WDs in a dynamic process such as tidal captures. Subsequently, GW radiation causes their orbits to continuously shrink, and detached BH-WD binaries appear as low-frequency GW sources that can be detected by space-borne GW detectors. Once WDs enter the tidal radius of BHs, they will be disrupted by BHs, and these systems become UCXBs. For a semi-detached BH-WD binary with $M_{\rm BH} =8~ M_{\odot}$ and $M_{\rm WD} =0.2~ M_{\odot}$, the estimated minimum GW frequency is $6.4~\rm mHz$ \citep{qin23}. Because the tidal radius of the BH satisfies \citep{hopm04}
\begin{equation}
R_{\rm t}=\left(\frac{M_{\rm BH}}{M_{\rm WD}}\right)^{1/3}R_{\rm WD}\propto M_{\rm WD}^{-2/3},
\end{equation}
the tidal radius will be smaller if the WD captured by the BH is more massive. According to the Keplerian third law, it would yield a relatively high minimum GW frequency. However, the maximum GW frequency is $5.6~\rm mHz$ for  BH UCXBs evolving from
the He star channel. Therefore, the GW frequency could be a probe to test the formation channel of BH UCXBs.
\begin{figure}
\centering
\includegraphics[angle=0,width=1.15\linewidth,trim={0 0 0 0},clip]{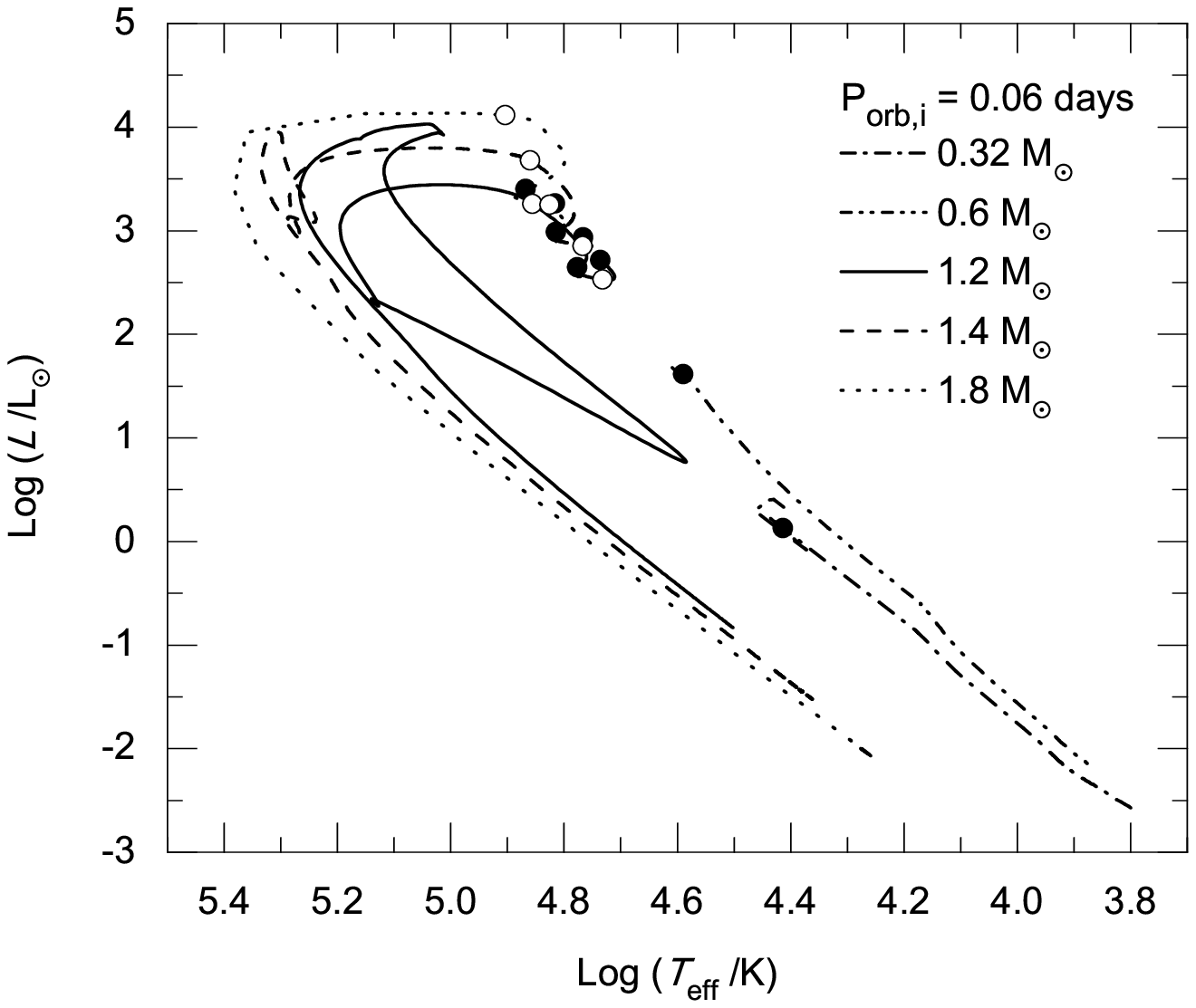}
\caption{Evolutionary tracks of several BH+He star binaries with different initial donor-star
masses and an initial orbital period of 0.06 days in the H-R diagram. The
solid and open circles represent the onset and end of mass transfer, respectively} \label{fig:orbmass}
\end{figure}

\subsection{Influence of tidal effects}
In the detailed binary evolution models, we ignore the tidal effects. In principle, the tidal coupling between the orbit and the He donor star can influence the orbital evolution of BH X-ray binaries. During the shrinkage stage of the orbit, the donor star spins up to corotate with the orbital rotation due to tidal coupling. This spin-up indirectly consumes the orbital angular momentum, extracting orbital angular momentum from the binary system at a rate of
\begin{equation}
\dot{J}_{\rm t}=-I\dot{\Omega},
\end{equation}
where $I$ is the momentum of inertia of the He star, $\dot{\Omega}$ is the derivative of the orbital angular velocity. Ignoring the influence of the mass transfer, the change rate of the orbital angular velocity $\Omega$ satisfies
\begin{equation}
\frac{\dot{\Omega}}{\Omega}\sim-3\frac{\dot{J}_{\rm gw}}{J},
\end{equation}
where $J$ and $\dot{J}_{\rm gw}$ are the total angular momentum of the system and its loss rate via gravitational radiation. Therefore, the ratio between $\dot{J}_{\rm t}$ and $\dot{J}_{\rm gw}$ is
\begin{equation}
\frac{\dot{J}_{\rm t}}{\dot{J}_{\rm gw}}\sim3\frac{I\Omega}{J}=3\frac{I}{\mu a^{2}},
\end{equation}
where $\mu=M_{\rm BH}M_{\rm He}/(M_{\rm BH}+M_{\rm He})$ is the reduced mass of the binary system.

Considering a BH binary with $M_{\rm BH}=8~M_{\odot}, M_{\rm He}=1~M_{\odot}$, and $P_{\rm orb}=0.1~\rm days$, we have $\mu a^{2}=3.0\times10^{55}~\rm g\,cm^{2}$. Since the radius of the He star $R_{\rm He}=0.212R_{\odot}(M_{\rm He}/M_{\odot})^{0.654}$ \citep{taur06}, the momentum of inertia of the He star can be estimated to be $I=0.4M_{\rm He}R_{\rm He}^{2}=1.7\times10^{53}~\rm g\,cm^{2}$. According to equation (11), the rate of angular momentum loss via the tidal effects is approximately two orders of magnitude smaller than that via gravitational radiation. Therefore, the influence of the tidal effects on the orbital evolution is trivial.

\subsection{Influence of BH Masses on the Parameter Space}
The binary population synthesis simulations found that the newborn BHs have a mass range of $5-16~M_{\odot}$ in BH binaries with normal-star companions, in which the BH-masses distribution emerges a peak at $\sim 7-8~M_{\odot}$ in the Model A of \cite{shao19}. In BH-He star X-ray binaries, the BH masses range from 5 to $20~M_{\odot}$, and are most likely to gather at $\sim 7-8~M_{\odot}$ in the Model A \citep{shao20}. Based on these statistical results, we adopt a constant initial BH mass of $8~M_{\odot}$ in the detailed binary evolution models.

For a same He star, a high BH mass naturally results in a high chirp mass, a high characteristic strain, and a long detection distance of BH UCXB-LISA sources. When $M_{\rm He,i}=0.6~M_{\odot}$ and $P_{\rm orb,i}=0.06~\rm days$, our calculations found that the maximum GW frequencies in the UCXB stage are 2.55, 2.75, and 2.91 mHz for $M_{\rm BH,i}=15, 8$, and $5~M_{\odot}$, respectively. During the evolution BH binaries, a high BH mass would produce an efficient angular momentum loss via GW radiation, and a high mass transfer rate. The rapid mass transfer from the less massive He star to the more massive BH drives the orbits of the systems to widen, resulting a relatively small maximum GW frequency. Therefore, there exist inverse correlation between the initial BH mass and the maximum GW frequency in BH UCXBs evolved from the He star channel.

When $M_{\rm BH,i}=5~M_{\odot}$, and $M_{\rm He,i}=1~M_{\odot}$, our simulations show that the initial orbital periods of BH+He star binaries that can evolve into BH UCXB-LISA sources are in the range of $0.04-0.08$ days, which is slightly narrower than that ($0.04-0.09$ days, see also Figure 5) in $M_{\rm BH,i}=8~M_{\odot}$. If the BH has a high initial mass of $15~M_{\odot}$, the range of initial orbital periods turns into $0.04-0.1$ days. Therefore, a high/low initial BH mass tends to widen/reduce the initial orbital period range. This tendency originates from the gravitational radiation is the dominant mechanism driving the orbit of BH+He star binaries to shrink. A high mass BH naturally produces a high rate of angular momentum loss, driving the BH+He star binaries with a slightly long period to evolve into BH UCXB-LISA sources. Furthermore, the initial He-star masses that can evolve into BH UCXB-LISA sources are in the range of $0.32-3.0~M_{\odot}$ and $0.32-2.9~M_{\odot}$ for $M_{\rm BH,i}=5$, and $15~M_{\odot}$, respectively. Therefore, the influence of the initial BH masses on the initial parameter space is trivial.

\subsection{Detectability of BH UCXBs as LISA sources}
We employ a rapid binary evolution code developed by \cite{hurl00,hurl02} to study the birthrate of BH UCXB-LISA sources in the Galaxy. Based on the binary population synthesis (BPS) approach, the primordial binary samples are produced in the way of Monte Carlo simulations. Subsequently, a sample of $1 \times10^{7}$ primordial binaries are evolved until the formation of BH+He star systems through the rapid binary evolution code. Similar to \cite{chen20}, the initial input parameters and basic assumptions in the BPS simulation are as follows: (1) All primordial stars are thought to be members of binary systems orbiting in circular orbits. (2) The primordial primary mass distribution arises from the initial mass function derived by \cite{mill79}, and it produces the secondary mass distribution by adopting a constant mass ratio ($0<q\leq1$) distribution $n(q)=1$. (3) The initial separations distribution is taken to be constant in ${\rm log} a$ for wide binaries with orbital periods longer than 100 yr, then changes into a uniform distribution for close binaries \citep{eggl89}. (4) The standard energy prescription proposed by is used to tackle the CE ejection process \citep{webb84}, in which the efficiency $\alpha_{\rm CE}$ of ejecting the envelope and the parameter $\lambda$ describing the stellar mass-density distribution are merged as a degenerate parameter $\alpha_{\rm CE}\lambda$. BH UCXB-LISA sources are thought to be produced if the parameters of the BH+He star systems are consistent with those of the progenitors of BH UCXB-LISA sources in Figure 5. The influence of the initial BH mass on the initial parameter space is trivial, thus its affect on the population synthesis simulations can be ignored. As a consequence, the criterion determined compact objects in the BPS simulation is just a BH, no matter whether their masses are $8~M_{\odot}$.

In Figure 9, we plot the evolution of the birthrates of BH UCXB-LISA sources evolving from the He star channel as a function of time when we take a constant star formation rate (SFR) of $5~M_{\odot}\, \rm yr^{-1}$ for Population I.  The Monte Carlo simulations predict the birthrates of BH UCXB-LISA sources in the Galaxy to be $R=2.2\times10^{-6}$, and $3.6\times10^{-7}~\rm yr^{-1}$ when the degenerate CE parameter $\alpha_{\rm CE}\lambda=1.5$, and 0.5 \citep[$\alpha_{\rm CE}$ is the CE ejection efficiency, and $\lambda$ is the stellar structure parameter,][]{webb84}, respectively. These two birthrates are $1-2$ orders of magnitude higher than that ($\sim3.9\times10^{-8}~\rm yr^{-1}$) estimated in the MS channel \citep{qin23}, however, are one order of magnitude lower than those ($3.1-11.9\times10^{-6}~\rm yr^{-1}$) predicted for NS UCXBs evolving from the He star channel \citep{wang21}.

According to Table 1, the mean detection timescale $\bigtriangleup t_{ \rm LISA,10}\approx 15~\rm  Myr$ for BH UCXB-LISA sources evolving from the He-star channel for a detection distance of 10 kpc. Therefore, we can estimate the detection number of BH UCXB-LISA sources formed by the He-star channel in the Galaxy to be $N=R \bigtriangleup t_{ \rm LISA,10}\approx 33$, and 5 as $\alpha_{\rm CE}\lambda=1.5$, and 0.5, respectively. For a low degenerate CE parameter $\alpha_{\rm CE}\lambda=0.5$, the detection number of BH UCXB-LISA sources evolving from the He-star channel is similar to that from the MS channel \citep[$\approx 4,$][]{qin23}. However, the detection number of BH UCXB-LISA source evolving from the He star channel is one order of magnitude higher than that from the MS channel for a conventional degenerate CE parameter $\alpha_{\rm CE}\lambda=1.5$.

\begin{figure}
\centering
\includegraphics[angle=270,width=1.0\linewidth,trim={0 0 0 0},clip]{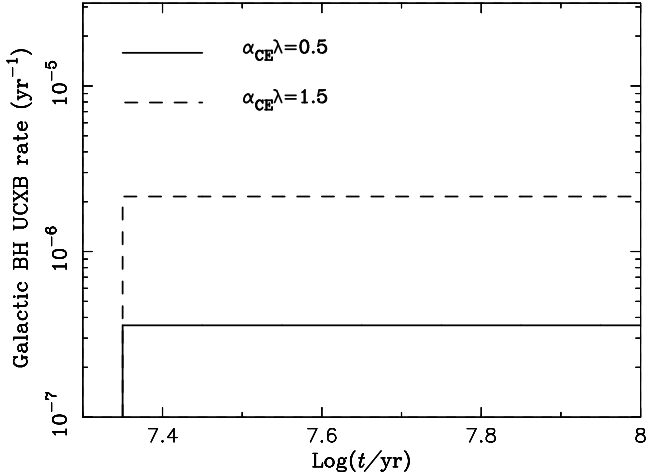}
\caption{Evolution of the birthrates of BH UCXB-LISA sources evolving from the He star channel as a function of time when we take a constant SFR of $5~M_{\odot}\, \rm yr^{-1}$ for Population I. The solid, and dashed curves represent the evolutionary tracks when the degenerate CE parameter $\alpha_{\rm CE}\lambda=0.5$, and 1.5, respectively.} \label{fig:orbmass}
\end{figure}
\section{Conclusion}
BH UCXBs can emit both low-frequency GW signals and X-ray emission, making them intriguing multi-messenger detection sources. In this paper, we employ the MESA code to simulate the formation and evolution of BH UCXBs evolving from the He star channel and diagnose their detectability in both X-ray and GW bands. Taking an initial BH mass of $M_{\rm BH,i}=8~M_{\odot}$, our main conclusions are summarized as follows:
\begin{enumerate}
\item The mass transfer of BH+He star binaries is sensitive to the masses of He stars. When $M_{\rm He,i} \geq 0.7~M_{\odot}$, those systems may experience two RLOFs. The mass-transfer rates in second RLOF stages can reach $\sim 10^{-6}~M_{\odot}\, \rm yr^{-1}$. Such a high mass-transfer rate causes their orbits to rapidly expand, thus some systems can not appear as UCXB-LISA sources at a distance of 10 kpc in the later stage of the second mass-transfer phase.

\item The mass-transfer rates of BH UCXBs evolving from the He star channel are much higher than those from the MS channel,  producing X-ray luminosities that are $2-5$ orders of magnitude higher than those from the MS channel. Due to the short initial orbital periods, the BH+He star systems become UCXBs at the beginning of the mass transfer and emit an X-ray luminosity of $\sim10^{38}~\rm erg\, s^{-1}$. The maximum X-ray luminosity in the first mass-transfer stage can exceed $10^{39}~\rm erg\, s^{-1}$. Those BH binaries with massive He stars experience a second RLOF, and produce a maximum X-ray luminosity of $3.5\times 10^{39}~\rm erg\, s^{-1}$, which exceeds the threshold luminosity of ULXs.

\item The shortest orbital period in the BH UCXB stage is 6 minutes, which corresponds to a GW frequency of 5.6 mHz. Because of short initial orbital periods, our simulated BH+He star binaries are already BH UCXB-LISA sources within a distance of 10 kpc (or 100 kpc) at the onset of the first mass transfer. BH X-ray binaries with massive He stars ($M_{\rm He,i}=1.8$, and $2.8~M_{\odot}$) will not be detected by LISA at a distance of 10 kpc in the later stage of the second mass-transfer phase due to a rapid orbital expansion. However, the continuous orbital shrinkage due to GW radiation causes the system with $M_{\rm He,i}=1.8~M_{\odot}$ to evolve toward LISA sources after the second mass transfer ceases, emitting GW signals with a frequency of up to 63.5 mHz.

\item Compared with NS UCXBs from the He star channel \citep{wang21}, the progenitors of BH UCXB-LISA sources have a slightly wide initial parameter space. The initial He-star masses and initial orbital periods of the progenitors of Galactic BH UCXB-LISA sources are in the range of $0.32-2.9~M_{\odot}$ and $0.02-0.19$ days, respectively. Meanwhile, nearly all systems in this parameter space can evolve into BH UCXB-LISA sources whose chirp masses can be accurately measured. However, only a tiny fraction of BH UCXB-LISA sources possess measured chirp masses for the MS channel \citep{qin23}.

\item In the He star channel, those BH binaries with a massive He star ($1.2-1.8~M_{\odot}$ ) can evolve into detached BH-WD binaries, which can not be achieved in the MS channel. Meanwhile, the effective temperatures of those WDs evolving from the He star channel are much higher than those of WD evolving from the MS channel because of the absence of a thin H envelope.

\item The Monte Carlo BPS simulations predict the birthrates of Galactic BH UCXB-LISA sources evolving from the He star channel to be $R=2.2\times10^{-6}$, and $3.6\times10^{-7}~\rm yr^{-1}$ when $\alpha_{\rm CE}\lambda=1.5$, and 0.5, respectively.
    In a 4-year LISA mission, the detection numbers of Galactic BH UCXB-LISA sources for the He-star channel are 33, and 5 for $\alpha_{\rm CE}\lambda=1.5$, and 0.5, respectively.

\item Compared with the MS channel, BH UCXB-LISA sources from the He star channel possess high X-ray luminosities and high possibility of detecting the chirp masses. Therefore, BH UCXB-LISA sources evolving from the He star channel are ideal multimessenger objects that deserve to be pursued in the X-ray and GW community.
\end{enumerate}

\acknowledgments {We cordially thank the anonymous referee for the detailed and insightful comments that improved this manuscript. This work was partly supported by the National Natural Science Foundation of China (under grant Nos. 12273014, 12225304, 12273105, 12373044, and 12203051), the Western Light Project of CAS (No. XBZG-ZDSYS-202117), the Youth Innovation Promotion Association CAS (No. 2021058), and the Natural Science Foundation (under grant
No. ZR2021MA013) of Shandong Province.}

\end{document}